\documentclass[prd,tightenlines,twocolumn]{revtex4}
\usepackage{amsmath}
\usepackage{amssymb}
\usepackage{graphicx}
\usepackage{bm}
\usepackage{enumerate}
\usepackage{color}
\newcommand{\be}{\begin{equation}}
\newcommand{\ee}{\end{equation}}
\newcommand{\bea}{\begin{eqnarray}}
\newcommand{\eea}{\end{eqnarray}}
\newcommand{\ba}{\begin{eqnarray}}
\newcommand{\ea}{\end{eqnarray}}

\begin{document}

\title{ New Regimes of Stringy (Holographic) Pomeron \\ and 
High Multiplicity $pp$ and $pA$ Collisions }
\author{ 
 Edward Shuryak and Ismail Zahed}

\affiliation{Department of Physics and Astronomy, \\ Stony Brook University,\\
Stony Brook, NY 11794, USA}

\date{\today}

\begin{abstract} 
   Holographic AdS/QCD models of the Pomeron unite a string-based description
  of hadronic reactions of the pre-QCD era with the perturbative BFKL approach.
  The specific version we will use due to Stoffers and Zahed \cite{Stoffers:2012zw,STOFFERS,ZAHED,Stoffers:2012ai,Basar:2012jb}, 
  is based on  a semiclassical quantization of a ``tube" (closed string exchange or open string virtual pair production)
 in its Euclidean formulation using the scalar Polyakov action.  This model has a number of phenomenologically successful results.
The periodicity of a coordinate around the tube allows the introduction of a
Matsubara time and therefore an effective temperature $T_{eff}$ on the string. We observe that at the LHC energies and for sufficiently small impact parameter,  $T_{eff}$
 approaches and even exceeds the Hagedorn temperature of the QCD strings.
    Based on studies of the stringy thermodynamics of pure gauge theories we  suggest
   that there should exist two new regimes of the Pomeron: the ``near-critical"  and the ``post-critical" ones.
  In the former one, string excitations  create  a high entropy ``string ball",
with high energy and entropy but small pressure/free energy. If heavy enough this ball becomes  a (dual) black hole
(BH).  As the intrinsic temperature of the string  exceeds  the Hagedorn temperature, the  ball  becomes a post-critical  explosive ``QGP ball".  The hydrodynamical explosion resulting from this scenario was predicted  \cite{Shuryak:2013ke} to have radial flow
exceeding that ever seen even in heavy ion collisions, which was recently confirmed by CMS and ALICE at LHC.
 We also discuss the elastic scattering profile, finding some hints for new phases in it, as well as two-particle correlations.
 \end{abstract}

\maketitle
\section{Introduction}
\subsection{The main ideas}

   Historically, the description of strong interactions has been shifting between
  an emphasis on perturbative and non-perturbative physics. This can be seen in
  the theory of hadronic collisions as well. The phenomenology of the 1960 and 1970 has
  revealed      Pomeron and Reggeon exchanges, which later -- due to Veneziano and others --  
were shown to be related with strings. The discovery of QCD gave rise to weak coupling approaches, instrumental
for hard processes. When theorists turned to hadronic collisions in the Regge kinematic
$s\gg |t| $ in the perturbative approach, they found the so called BFKL Pomeron \cite{BFKL}, 
through the re-summation of gluonic  ladders. 

After the discovery of the AdS/CFT correspondence, the last decade 
saw the rapid developments of holographic models, collectively called AdS/QCD, which try to unify both
weak and strong coupling regimes within the same framework. 
Holography adds an extra dimension of space, identified with the ``scale" in the sense of the renormalization group.
The  ultraviolet (UV) end of this space is at weak coupling and  large momentum transfer $|t| \gg \Lambda^2_{QCD}$, 
while the infrared (IR) part is at strong coupling appropriate to  small 
 $  |t| \sim 1\, {\rm GeV}^2$ in the typical hadronic collisions. In this work we will use a particular
version of such a model developed by Stoffers, Zahed and others \cite{Stoffers:2012zw,STOFFERS,ZAHED,Stoffers:2012ai,Basar:2012jb} and based on scalar Polyakov strings propagating in the 5-dimensional holographic space. A historical evolution of the pomeron in
holography can be found in a number of references within the past decade~\cite{SIN,Janik:2000pp,Janik:2001sc,GRAVITON,BLACK,Hatta:2007cs,Albacete:2008ze} .

The understanding of the dynamics behind  Pomerons and Reggeons still remains a challenging task. Traditionally
 models have been judged by their predictions on a rather limited number of observables,
such as  the dependence of the total and elastic cross sections on $s$ and diffraction, related with
certain fluctuations in the system.  A 
new turn of events has taken place at the beginning of the LHC operation which has allowed to trigger on
very high multiplicity  events
\cite{Khachatryan:2010gv,CMS:2012qk,Abelev:2012cya,ATLAS,Adare:2013piz}. 
These events bring about novel issues related to strong fluctuations in
the collision system.

Before we delve into the specifics of our analysis, let us identify our main idea. In the macroscopic (thermodynamical) context
it is well known 
that the perturbative quark-gluon phase  and the non-perturbative confining (stringy) phase are related by a first order phase transition (for $N_c>2$ which we imply here). We argue that the same should be true for high energy scattering, as a function of the impact parameter, with all three regimes present in the scattering amplitude.

 It is a well known fact, that explaining confinement starting from gluons is an extremely difficult (not yet completed) task. On the other hand, going in the opposite direction -- from strings into the perturbative phase -- is easier, and it was in fact qualitatively understood long ago.
In  the stringy approach an explanation is in terms of the so called Hagedorn phenomenon \cite{Hagedorn}.  Strings have exponentially rising density of states, as first noted by Fubini and Veneziano
\cite{FV}. The explicit expression for the density of states $d(n)$ appeared in Huang and Weinberg \cite{HV}, with cosmological consequences a la Hagedorn. A decade later, after the discovery of QCD and its formulation on the lattice, this fact re-surfaced again in finite-temperature QCD, through the work of Polyakov and Susskind~\cite{Polyakov:1979gp,Susskind:1979up}. 

\begin{figure}[t]
  \begin{center}
  \includegraphics[width=7cm]{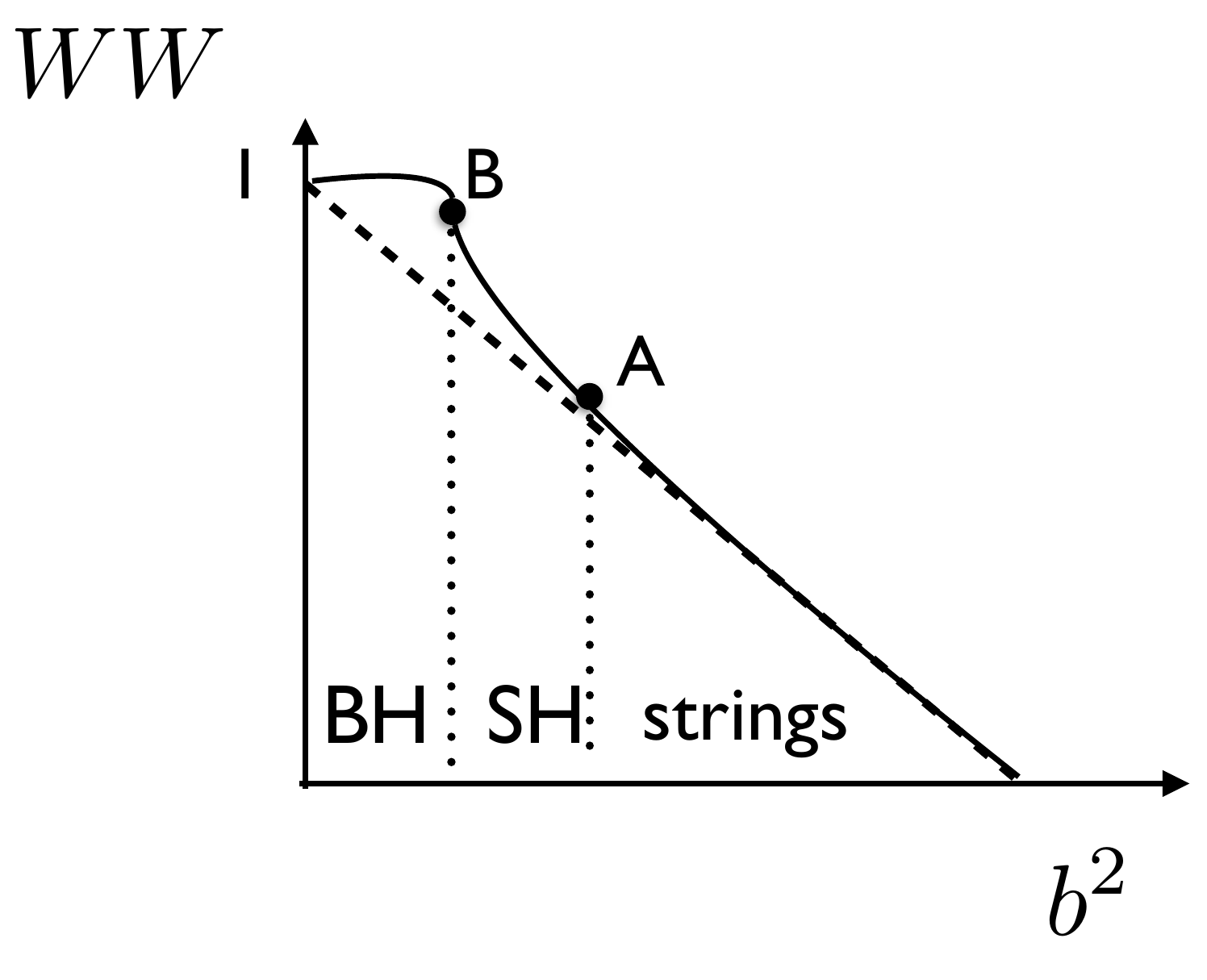}
  \caption{(Color online) Schematic representation of the log of the dipole-dipole scattering amplitude as a function of 
  the squared impact parameter
  ${\bf b}^2$. The dashed line is the Gaussian-shaped string amplitude. 
  The solid line represents the result, in three different regimes.
  For an explanation of BH (Black Hole) and SH (String Hole) see text.}
  \label{fig_amplitude}
  \end{center}
\end{figure}

Based on the analogy to thermodynamics of the glue (for some technical details and references see section \ref{sec_fluct})  we will argue that in high energy collisions  the excitations of the  exchanged non-perturbative objects
(two open strings or a closed string)  should also proceed subsequently through {\em three distinct stages},
as one proceeds from more peripheral to more central collisions:
\\
\\
\indent{{\bf 1.}~A ``cold"  or sub-critical regime, with low string excitations, that  generates a Pomeron with
a Gaussian profile;}
\\
\\
\indent{{\bf 2.}~A ``near-critical"  regime, in which the exchanged string effectively decreases its tension due to the Hagedorn phenomenon,
 but increases its energy and entropy and turns to a ``string ball". With the inclusion of self-interaction, the excited string
 is prone to implosion. It 
reduces its size and transmutes to a black hole. The corresponding transitory object is called a $string-hole$ or SH for short;}
\\
\\
\indent{{\bf 3.}~A super-critical or ``explosive" regime, in which the string  becomes a black hole, BH for short,
corresponding effectively to string breaking and the deconfined QGP phase.  
The Hawking radiation creates a perturbative thermal state, which generates sufficient pressure and leads to
 hydrodynamical explosion.}
\\
\\

\begin{figure}[b]
  \begin{center}
  \includegraphics[width=8cm]{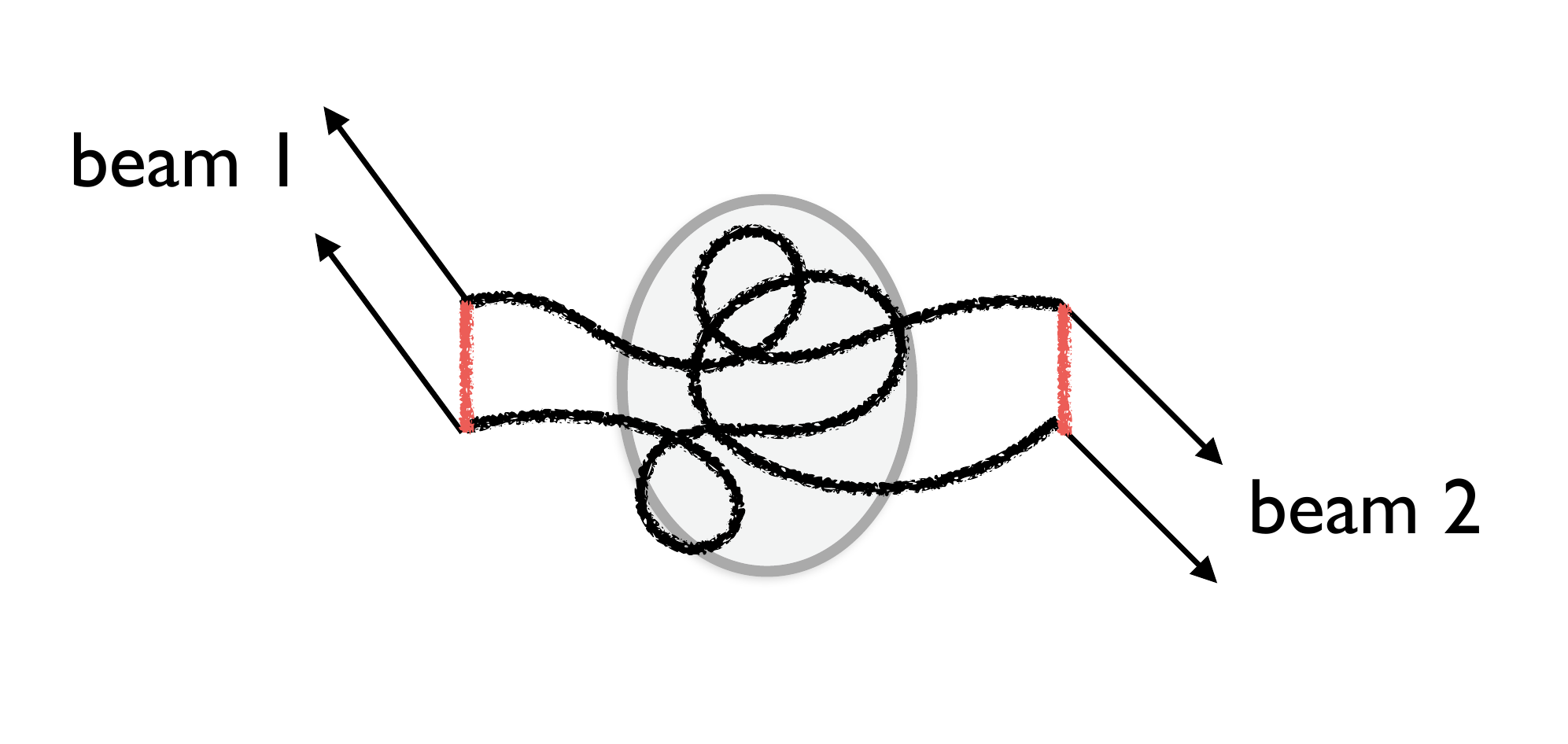}
  \caption{(Color on-line)  A sketch of a string configuration at $t=0$, as it appears from the under-the-barrier Euclidean domain.
  The small size dipoles are an approximation to colliding protons. At $t=0$ they are separated by the transverse distance ${\bf b}$, the impact parameter. They move in the direction shown by two arrows later. The gray shaded sphere indicates a gravitational trapped surface. }
  \label{fig_ball}
  \end{center}
\end{figure}

A sketch of the scattering amplitude versus the (squared) impact parameter ${\bf b}^2$, displaying all three regimes, is 
shown in Fig.~\ref{fig_amplitude}. The details of the plot as well as the approximations used and the objects under considerations will be clarified as we proceed.
   At this point, let us just supply a sketch of the string ball, in Fig.~\ref{fig_ball}. If heavy enough, its (effective) gravity may generate an
   effective trapped surface,
   shown by a gray circle. While we do not investigate the existence and properties of the trapped surfaces in this work,
   we would like to mention two works \cite{Lin:2009pn,Gubser:2009sx} which did study those,  in different but related context, and concluded    that the trapped surface suddenly disappears  above a critical value of the impact parameter.

   Returning to recent developments, we note that the current LHC experiments provide 
  high luminosity and high-rate detectors, capable to detect and study very low probability fluctuations of the system. 
In the first LHC $pp$ run, the CMS collaboration~\cite{Khachatryan:2010gv}  has used  
this opportunity and triggered on events with high multiplicity. This was followed by  similar (but much less expensive)
triggered studies in 
 pPb \cite{CMS:2012qk} .  Multiple studies to follow -- including experimental~\cite{Abelev:2012cya,ATLAS,Adare:2013piz}
  and theoretical papers associated those observations with the production of a small-size hot fireball made of a
  Quark-Gluon Plasma (QGP) , that explodes hydro-dynamically. Those recent papers include 
  ours \cite{Shuryak:2013ke}, in which we predicted that the radial flow in high multiplicity $pp$ and $pA$ events should be even stronger than in $AA$  collisions. Radial flow has been recently observed by CMS and ALICE
  via spectra of identified particles, confirming our  theoretical prediction.
 
  

The paper is structured as follows: Since we aim at rather different readers, from
heavy ion experimentalists to string theorists, we provide 
two more subsections of the Introduction
containing a brief introduction to the Pomeron phenomenology and its stringy description \ref{sec_pomerons},
as well as  the thermodynamics of the glue \ref{sec_thermo}. (Experts obviously may omit some of that.)
The main body of the paper starts in Section II from a review of glueball Regge trajectories  \ref{sec_glueballs}
and their relation to particle correlations. We emphasize the role of
correlation measurements for finding ``clustering" of hadrons, related in the Regge language with 
the exchange of the excited ("daughter") Pomerons. In section
\ref{sec_SZ_model} we introduce the physical setting and 
the main results of the SZ Pomeron model, including  its weak coupling limit \ref{sec_BFKL}
and daughter trajectories \ref{sec_reggions_SZ}. 

The core of the paper is section  \ref{sec_fluct} devoted to quantum fluctuations of the exchanged strings. 
In spite of the fact that we are dealing with a zero temperature scattering amplitude, in sub-section \ref{sec_entropy} we explain that string excitations naturally
have a thermodynamical description including temperature and entropy. Those take  the central stage as we discuss in section \ref{sec_near_critical} 
the  near-critical regime and in section \ref{sec_supercritical} the super-critical one.
The main ideas happen to be well developed in the string theory literature. They include
the transition to a black hole and a ``thermal scalar formalism".
Section \ref{sec_observables} discusses observable consequences of the scenario. Sub-section 
\ref{sec_elastic} is devoted to the elastic scattering amplitude. We compare our predictions
with a parameterization of the data, and show that it contains evidences of the change of behavior consistent with our
interpretation. In subsection \ref{sec_inelastic} we discuss predictions for a cluster
produced in high multiplicity inelastic collision, in particular its $t$-channel description in terms of the Pomeron daughter exchange. 
The remainder of the paper contains additional theoretical considerations, further elucidating the connection between 
  a string-ball and  a black hole, see section \ref{sec_horizon}. One result is 
the value of the ``string viscosity", and also a discussion of the Hawking radiation \ref{sec_Hawking}.
In our final discussion section we provide a summary of the results \ref{sec_summary}.

\subsection{  Pomerons, Reggeons and QCD strings} \label{sec_pomerons}

The Pomeron is an effective object corresponding to the   the highest  Regge trajectory $\alpha(t)$
and dominating the high energy cross sections at small $|t|\ll s$
\be 
{d\sigma \over dt} \approx \left( \frac{s}{s_0}\right)^{\alpha(t)-1 } \approx e^{{\rm ln}(s)(\alpha(0)-1)+\alpha ' t} 
\ee
 Originally  Pomeranchuk and Gribov \cite{Gribov:1962fw} suggested
a universal pole with vacuum quantum numbers and the intercept $\alpha_P(0)-1=0$,  corresponding to 
an asymptotically constant cross sections. The discovery of slowly rising  cross sections  $\sigma_{hh}(s) $   
 led to the so called ``supercritical soft Pomeron"  with   $\alpha_P(0)-1\approx 0.08$. 
  Regge trajectories with various quantum numbers are subdominant and the corresponding cross sections are
 decreasing powers of $s$. For  example the leading $\rho$ meson trajectory has $\alpha_\rho(0)-1\approx -0.5$. 
 The glueball (Pomeron daughters) excitations  have even smaller intercepts  $\alpha_{Pn}(0)<0$ to be discussed  below.

Diffractive processes with large rapidity gaps were described in terms of
 interacting Pomerons and Reggeons, which led Gribov~\cite{GRIB} and others to formulate some effective Reggeon Field 
Theory.  Important for the applications to diffractive and inelastic processes are the so called
AGK cutting rules~\cite{AGK}.  At large $\sqrt{s}$  it is a non-relativistic-like
field theory of interacting particles  (wee partons) diffusing in transverse  dimensions, with
the rapidity playing the role of time. 
The concept of  Gribov diffusion explains why the transverse size of a hadron 
grows with ${\rm ln}(s/s_0)$ (the rapidity or ``time" interval),  as observed in $pp$ and $p\overline{p}$  scattering. 
Pomerons interact but with a small triple-Pomeron vertex.  For recent Pomeron parameters and a fit to 
the LHC data on cross sections and multiplicities see e.g. \cite{Levin}. We note 
 that the intercept for the ``input Pomeron" used there is  $\alpha_{\bf P}(0)-1\approx 0.25$, amusingly
 similar to our starting Pomeron in flat space.

   In the pre-QCD period, 
the discovery of many s-channel resonances with conjectured t-channel Reggeon exchanges
led Veneziano  to the famed amplitude 
for the scattering of 2 scalars possessing planar duality between the s- and t-channel poles \cite{Veneziano:1968yb}.
This observation, was soon
 generalized to the scattering of N scalars and the dual resonance model. The
various attempts to  understand the meaning of these formulae led to  the idea   
of quantum strings rather than particles, underlying the string interactions at strong coupling. (This in turn led to the discussion 
about the internal consistency of the string formulation and to  the  fundamental superstring theory.)

Gribov partonic description of the Pomeron and its transverse diffusion follows from QCD
at weak coupling by re-summing rapidity ordered gluon or BFKL   ladders~\cite{BFKL}. 
At large $\sqrt{s}$ collinear gluon bremstralung is large even at weak coupling
and requires re-summation.  The 1-loop BFKL re-summed ladders lead to a perturbative
Pomeron with a large intercept and zero slope. 
(A formidable 2-loop calculation of the intercept of the
QCD perturbative Pomeron raises, once again, the issue of convergence of the perturbative series
at such $t$.) 

\begin{figure}[t]
  \begin{center}
  \includegraphics[width=6cm]{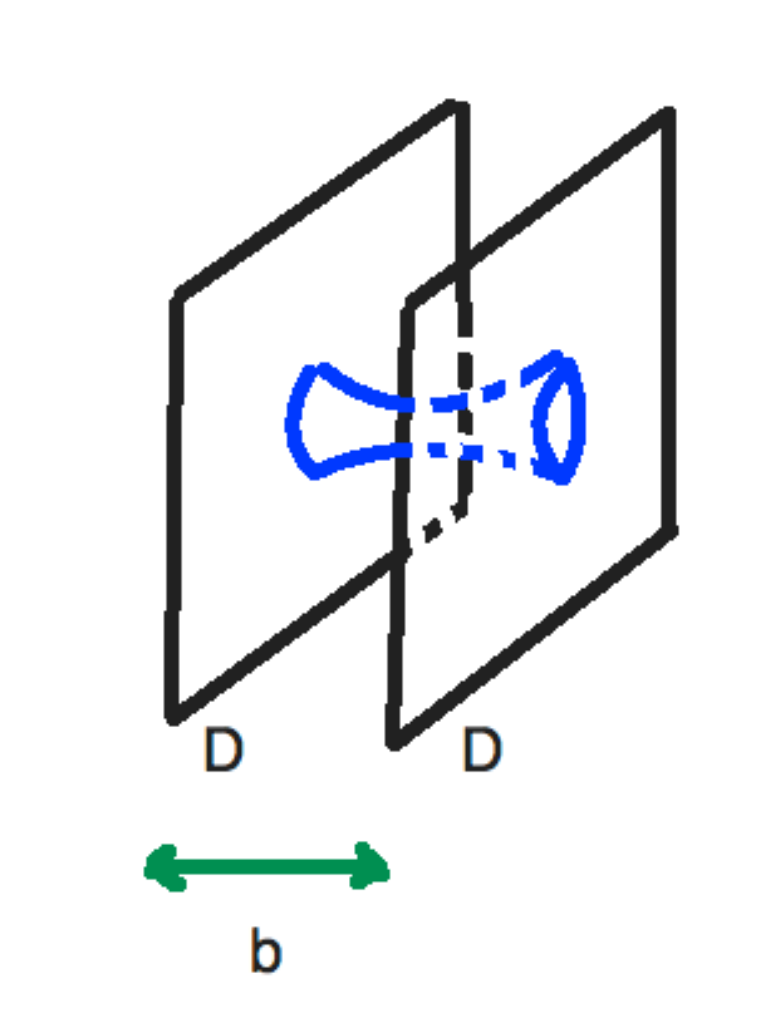}
  \caption{(Color on-line) Dipole-dipole scattering due to closed string exchange. The impact parameter  ${\bf b}$
  is the dipole transverse separation.}
  \label{fig_DD}
  \end{center}
\end{figure}

The  t'Hooft large $N_c$ with $\lambda=g^2 N_c$ fixed, and its planar diagrammatics 
led to speculations that  at strong coupling perturbative ``fish-net" diagrams generate a surface
\cite{Greensite:1984sb}. The   discovery of  string-gravity duality in the AdS/CFT holographic context  
\cite{Maldacena:1998im} makes the speculation more quantitative for certain gauge theories,  unfortunately not (yet) for confined QCD.

 A schematic picture of the (color) dipole-dipole scattering via a tubed-shaped surface exchange is shown
 in Fig.\ref{fig_DD}. It can be alternatively viewed as an exchange of a closed string glueball state,
 or a virtual production of a pair of open strings, which later annihilate each other. 
  The derivation of the
elastic and inelastic amplitudes generated by surface exchanges 
 were  addressed using bosonic variational surfaces~\cite{SIN,Janik:2000pp,Janik:2001sc},
see also 
 a black-disk model~\cite{BLACK}. 

(It has  been realized that
 in pure AdS with ${\cal N}$=4 supersymmetry and conformal symmetry
 the dominant scattering mechanism should be associated with a 
spin-2 graviton exchange ~\cite{GRAVITON}. This is not the case in the setting we have.
In particular, the main contribution is to the real part of the scattering amplitude, not related with inelastic
events we discuss.)  
 
 To put things in perspective it is worth reviewing the  phenomenology of the 
elastic $pp$ cross section $d\sigma/dt$.
Its behavior is studied experimentally all the way to LHC energies,  
see especially the results of the TOTEM collaboration at $\sqrt{s}=7\, {\rm TeV}$ in
\cite{Antchev:2011zz}.  In short there is a very accurate exponential  
$e^{\alpha^\prime t}$ behavior at small $|t|$, for several decades, followed
by a dip at $|t|=0.53\, {\rm GeV}$ and then a power-like tail $|t|^p$ with $p\approx 7.8$.  
 A single dip means that the imaginary part of the amplitude changes sign  once. 

\begin{figure}[t]
  \begin{center}
  \includegraphics[width=6cm]{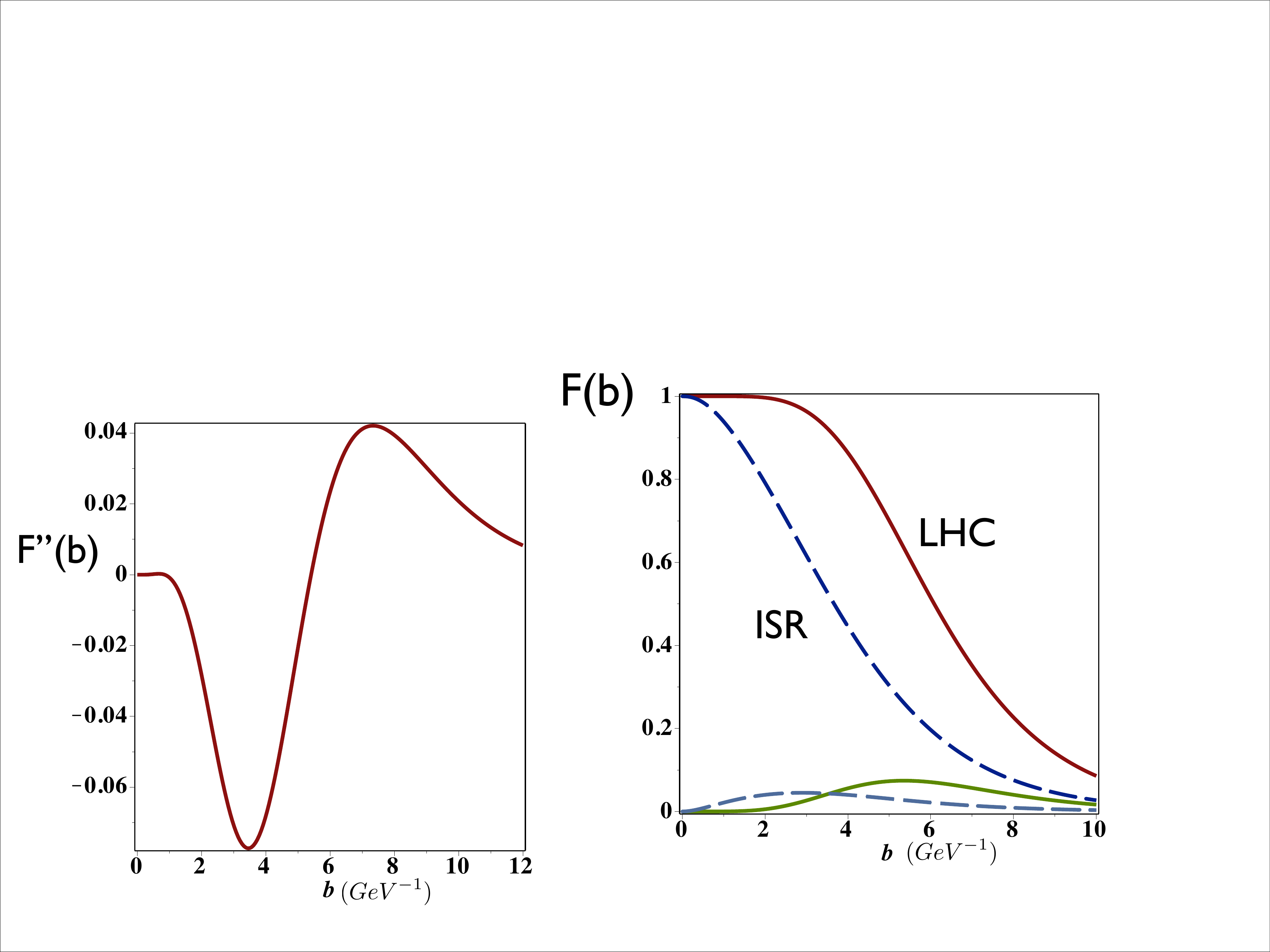} \\  \includegraphics[width=6cm]{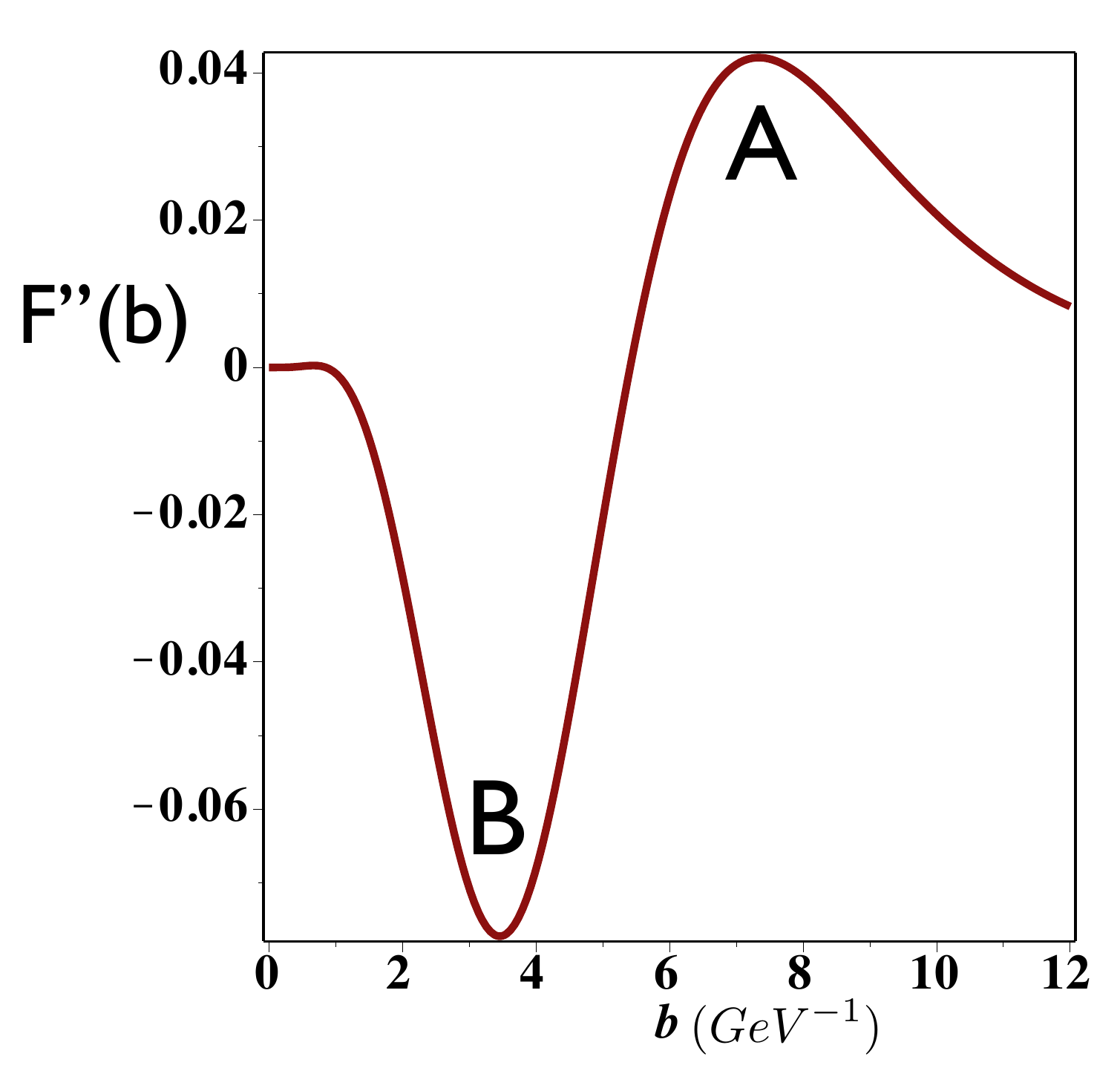}
  \caption{(Color online) The upper figure shows the imaginary (upper) and real (down) parts of the profile function $F(s,b)$ versus ${\bf b} ({\rm GeV}^{-1})$ 
  for $\sqrt{s}=7\, {\rm TeV}$ (solid) and $\sqrt{s}=63\, {\rm GeV}$ (dashed). The lower plot shows the second derivative over $b$   for $\sqrt{s}=7\, {\rm TeV}$ . Two maxima correspond to the same points $A,B$ as in the sketch in Fig.~\ref{fig_amplitude}.}
  \label{fig_profile}
  \end{center}
\end{figure}

 It is standard to use the impact parameter representation of the scattering amplitude, 
 connected with the momentum transfer via a Bessel transform 
 \be {\cal T}(s,q)= s \int_0^\infty d{\bf b}\,{\bf b} \,J_0({\bf b} q) \,{F}(s,{\bf b})  \label{eqn_profile}\ee
where $t=-q^2$ and $F(s,{\bf b})$  is the so called scattering profile. 
Since each set of data is taken only at some interval of $t$,
their direct Bessel-transform to coordinate space always include extrapolations. Instead of doing it numerically with data, one can do it instead analytically, with available parameterizations. Being a function of two variables -- $s,t$ -- it can be parameterized in multiple ways, and there is no shortage of models which can fit it.
An example  is the Bourrely-Soffer-Wu (BSW) model~\cite{Bourrely:2012hp}, see their expressions (13-15). These profiles are plotted in Fig.~\ref{fig_profile} for $pp$ collisions, at LHC and ISR energies.

 While the lower ISR energies have near-Gaussian shape, the LHC ones  display
three regions: ({\bf i}) a nearly horizontal plateau, ({\bf ii}) a relatively rapid turn downward,
and ({\bf iii}) an exponential  tail~\footnote{In this model as $exp(-m_1 {\bf b}), m_1=0.577\, {\rm GeV}$, 
perhaps representing $\sigma$ exchange.}.   
In order to see the boundaries of such three region more clearly, we also plotted in the lower 
plot of  Fig.~\ref{fig_profile}  the second derivative of the profile
function $F(s,{\bf b})$, at LHC energy. One can clearly see a positive and negative peak, indicating the ``turning points" of the profile.
We will argue below that these three regimes -- as a function of ${\bf b}$--  correspond
to the three dynamical regimes of a stringy Pomeron discussed in this work.

\subsection{Glueball Regge trajectories} \label{sec_glueballs}
   
   Nowhere in this paper the presence of quarks -- as fundamental color charges -- in QCD would be important, as
   all objects discussed are made of glue.  Of course,
 quarks  lead to string breaking (into mesons). However, this is  a tunneling and rather suppressed phenomenon,
 happening later in the process, after the system is out of its initial Euclidean phase.   
   
   Therefore in this paper we completely abstract ourselves from the existence of quarks (and
   quark-related states, the corresponding Regge trajectories etc.) and
   discuss only the physical states of pure gauge theory, the glueballs.
The glueball spectroscopy on the lattice is well developed, see  e.g. \cite{Morningstar:1999rf,Harvey},
but it is not widely known, so we will briefly review it.

\begin{figure}[t]
\includegraphics[width=7cm]{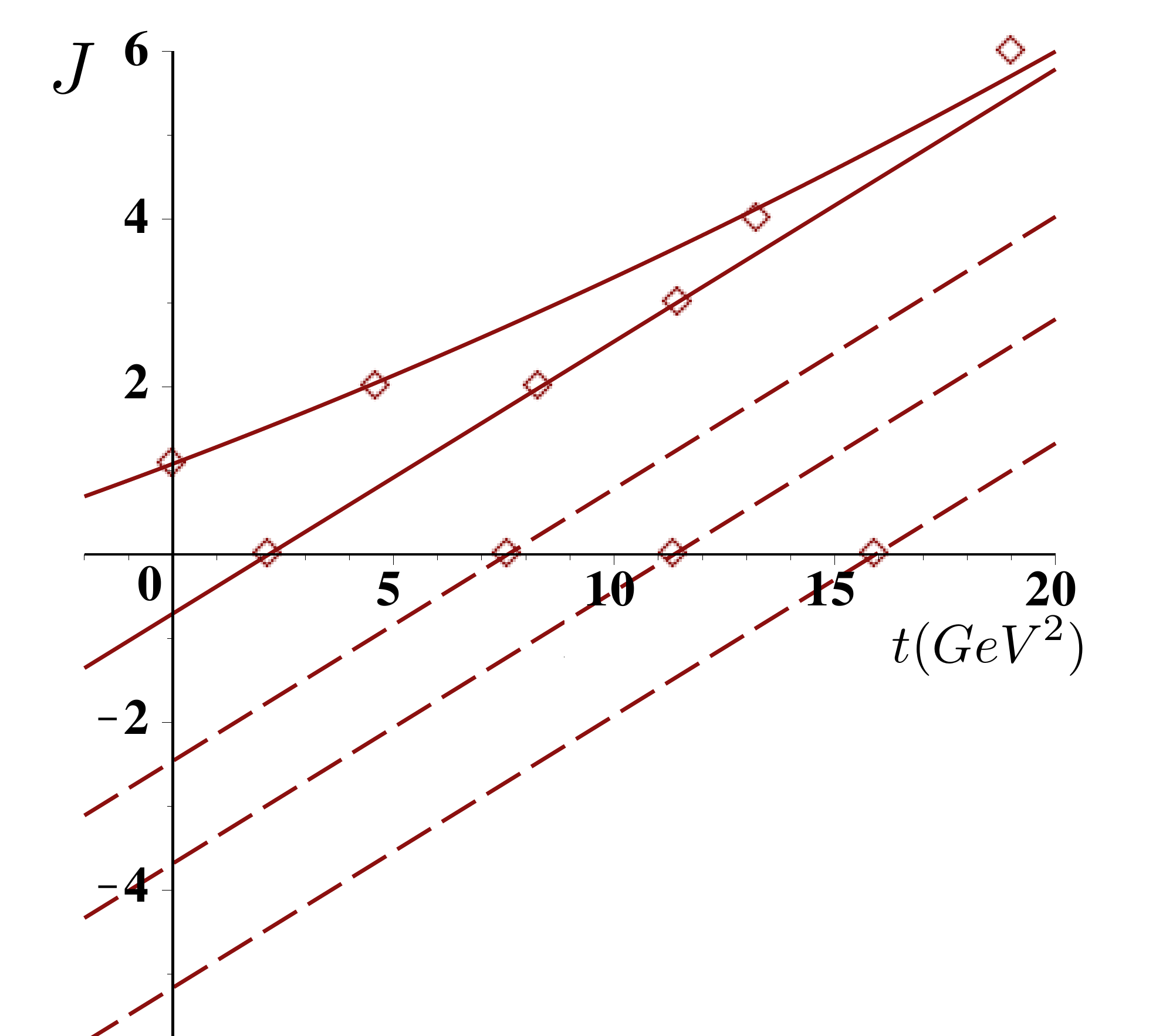}
  \caption{(Color online) Glueball Regge trajectories from lattice~\cite{Harvey}. }
  \label{glueballs}
\end{figure}

 In Fig.~\ref{glueballs} we display a compilation
of all $J^{PC}=J^{++}$ states defined in  the lattice simulations \cite{Harvey}.
Before we come to our main issue -- glueball Regge trajectories, a general comment is in order.
The lowest states -- which can be made of 2 gluons -- are scalar $0^{++}$ and tensor $2^{++}$ ones.
The forces mediated by those are both of attractive nature. (Those are in a way  the ``holographic images"
of the bulk graviton and dilaton of AdS/QCD, which are massive in the presence of the wall.)

There are several 
Regge trajectories associated to these states. The upper one includes four states, 
the Pomeron  and the $J^{++}=2^{++}, 4^{++}, 6^{++}$ states.
Its quadratic fit is 
\be J= \alpha(0) + \alpha' (0) M^2 +{\alpha''(0) \over 2} M^4 \ee
 $$\alpha'(0)=0.92/M^2_{2++},\,\,\,\,\alpha''(0)= 0.05/M^4_{2++}$$ using units of $M_{2++}=2.15\, {\rm GeV}$.
Its continuation to negative $t=- M^2$ is separately observable in scattering experiments. 

  The ``first daughter'' trajectory, consisting of three states $J^{++}=0^{++}, {2^*}^{++},3^{++}$, seems to be quite linear
with a negative intercept. Using the ``input Pomeron" \cite{Levin} 
one finds the intercept gap
\be \Delta \alpha_1=\alpha_P(0)-\alpha_{D1}(0)\approx 2.0  \label{daut1}    \ee
The next three daughter trajectories  (also indicated on the plot  by the dashed lines) have   
 only one -- the scalar --  excited glueball  in ~\cite{Harvey}, so in the plot we had to assume that all daughters  share the same slope (of course this needs  not be generally be correct). The second gap
 \be \Delta \alpha_2=\alpha_P(0)-\alpha_{D2}(0)\approx 4  \label{daut2} \ee
which is in overall agreement with the holographic result (\ref{T3}) below,
with the gaps 2 and 4, respectively.

The difference in slopes $\alpha'_{D1} > \alpha'_P$, observed in the glueball spectra 
is not predicted by the string models in flat space.
Physically this difference means that the states
of the daughter trajectories have larger spatial size than the Pomeron one.  Since the second daughter trajectory corresponds
to even higher excitations, their size and thus their slope $\alpha'_{D2}$ is perhaps also larger than $\alpha'_{D1} $. Thus
the gap between the intercepts $\Delta \alpha_2$ is perhaps  larger than the estimate above.

%
   

  As the number of states with momentum $J$ is $J(J+1)$ and $M_J\sim \sqrt{J}$ one might think that the density of states grows  as a power of the mass. However,   this is not so.  The number of stringy excitations 
 grows with the mass  {\em exponentially}. 
Thus, on one hand the states are on near-straight and approximately equidistant Regge trajectories. On the other hand,
the number of states grow exponentially. The resolution
of these seemingly contradicting statements lies in the fact that the daughter Regge trajectories
must be {\em multiply degenerate} (which is not shown on the figure, of course, as only special quantum numbers are selected).  
The high degeneracy $d(n)$ of the daughter trajectories with $n>0$ will be discussed in what follows. 

 \subsection{QCD strings and thermodynamics of the glue}  \label{sec_thermo}
   
   The most obvious and well known manifestation of the existence of the QCD strings is the approximate linear
   potential at large distances 
  
   \be V(r) \approx\sigma_T r 
   \ee 
   between fundamental color charges. 
    Stringy excitations manifest themselves in corrections to the linear potential, starting with the 
    famed Luscher term ${\cal O}(1/r)$ and its subleading corrections. Excitations of a string with particular
 quantum numbers have also been carried on the  lattice. For a  review discussing lattice results and their effective stringy description at various  $N_c$ see   \cite{Teper:2009uf}. In short, the lattice results indicate that 
the Nambu-Goto action -- tension times the area of the string worldvolume --
successfully describes all of those data.

Here, we mention an important theoretical result derived by Arvis \cite{Arvis:1983fp}, whereby 
the {\em re-summed} potential induced by the fluctuations of the  Nambu-Goto string resulted
 in the famous square root form

\be V(r)= \sqrt{ {r^2 \over (2\pi\alpha')^2} - { D_\perp \over 24 \alpha'}} \ee
Its expansion generates the so called universal Luscher terms mentioned above.

   In absolute values  the string tension 
        \be \sigma_T \approx (0.42\, {\rm GeV})^2 \label{eqn_sigma_T} \ee
        sets up the basic string units
     \be 2 \pi \sigma_T=  {1 \over \alpha'}={1 \over l_s^2} \ee
     Furthermore, following lattice conventions, we will also use it to define ``GeV" in all other confining theories, 
     including $SU(N_c)$ gluo-dynamics.

   Lattice simulations of gauge theories at finite temperatures and $N_c>2$  display 
   a first order transition $T_c\approx .27\, GeV$. (For details, such as the $N_c$ dependence of the 
   critical temperature $T_c$ and the latent heat see~\cite{Teper:2009uf}.)
   The thermodynamics of the glue at $T<T_c$ is very specific. Since the masses of 
   the glueballs (discussed above) are numerically large compared to $T$,
   they make an extremely dilute gas.  But the strings have so many states that 
   the excitations happen to originate  from the more massive states with an exponentially 
   rising degeneracy.

 As emphasized by Hagedorn~\cite{Hagedorn}, systems with exponentially growing density of states have 
 very peculiar thermodynamics, e.g. the thermal partition sum 
 \be 
 Z(T)=\int dE \,e^{E/T_H}\,e^{-E/T}\label{BOLTZ}
 \ee  
  diverges as $T\rightarrow T_H$,
 known 
 as the  Hagedorn temperature. Historically, Hagedorn  originally had a different picture of hadrons, as bags within
 bags in the bootstrap sense, not strings.   Hagedorn originally
  concluded that there exists a fundamental upper bound on temperatures,
  as such systems can reach infinite energy density with $T\rightarrow T_H$. The emergence
   of QCD in the seventies   and the development of the theory of the Quark Gluon Plasma showed
   that the Hagedorn phenomenon indicates a phase transition. Dedicated lattice studies  \cite{Bringoltz:2005xx}
   have shown that the Hagedrorn temperature is above the critical temperatures, namely  
   \be {T_H \over T_c}\approx 1.11 
   \label{THTc} \ee

   In the ``Hagedorn regime" at $T$ close to $T_H$ both the energy
  and entropy $S={\rm ln}N(L)$ are large, but in the free energy $F=E-TS$
 the two terms cancel out, causing $F$ to remain small. Since $F=-pV$, the string in the Hagedorn regime carries small pressure
 and does not explode.
 (Below we will further argue that near-critical strings should rather implode, due to their attractve self-interaction.)
 
   The simplest derivation of $T_H$ comes from ``coarse lattice" estimate by Polyakov.  
Imagine a d-dimensional lattice 
   with spacing $a\sim  l_s$ and draw all possible 
 strings of length $L/a$ making all possible turns (except going backward) at each site, that is
 \be
  N(E) \approx (2d -1)^{L/a}= e^{E(L)/T_H}
 \ee
 where in the last term we changed length into energy using the string tension $E(L)=\sigma_T L$ 
 This leads to
 \be 
 T_H= {\sigma_T  a \over {\rm ln}(2d-1)} 
 \ee
but in practice this is used  to estimate $a$ rather than $T_H$.
   
  Continuum strings lead to the expression
   \be (T^{QCD}_H)^2={3 \over D_\perp} {\sigma_T \over 2\pi} \approx  (0.176 \, {\rm GeV})^2 \label{THH}\ee
  which is indeed close to the critical temperature of the QCD deconfinement-chiral restoration transition.
In gluodynamics without quarks
    there are no mesons and baryons containing fundamental strings.
   Glueballs are made  of closed or double strings.
     The double string tension $2\sigma_T$ of such strings leads to a modified Hagedorn temperature 
     \be T^{YM}_H=  \sqrt{2} T^{QCD}_H \approx 0.237 \, {\rm GeV} \label{eqn_tilde_T_H}  \ee
     which  indeed approaches (but not  matches)   the lattice value mentioned above, namely
   $T_H^{YM}\approx 0.3\, {\rm GeV}$. 
   
%

\begin{figure}[t]
\begin{center}
\includegraphics[width=7cm]{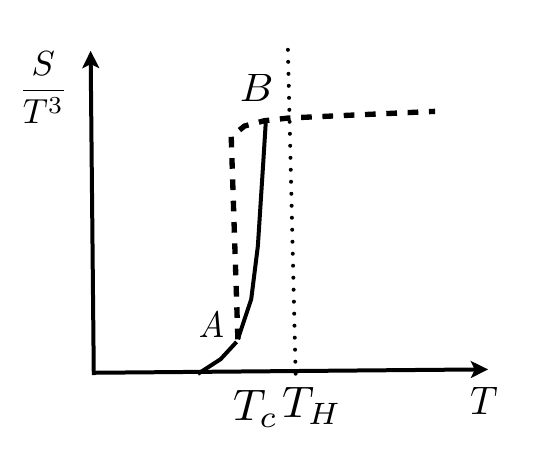}
\caption{(Color on-line) Schematic temperature dependence of the entropy density. The dashed line represents equilibrium
gluodynamics with a first order transition at $T=T_c$. The solid line between points $A$ and $B$
represents the expected behavior of a single string approaching its Hagedorn temperature $T_H$. 
The points $A$ and $B$, separating the intermediate phase, are in correspondence with our notations in Fig.\ref{fig_amplitude}.
}
\label{fig_entr}
\end{center}
\end{figure}

As the string density  gets  large enough, 
    the Hagedorn regime  ends  at the point $B$ of Fig.\ref{fig_entr}.
    This happens when  the energy (entropy) densities become as large as
 \be  
 {\epsilon \over T_H^4}\sim {s \over T_H^3} \sim  N_c^2 \label{condition_Nc}
 \ee 
 In this regime, the number of stringy degrees of freedom become higher than in the gluon gas and
 the latter becomes the preferable phase.  Thus, 
  a second qualitative change happens: the super-critical state  is the {\em deconfined} or  QGP
  phase. It can, however, still be described in a stringy language, as we will discuss below.

\section{The Holographic Pomeron}

\subsection{The SZ model} \label{sec_SZ_model}



The holographic approach used in the SZ model 
is known as  the ``bottom-up" one.   The holographic direction playing the role of the renormalization group scale,
describing in particular the sizes of the through-going dipoles.
There is  a large $N_c$ parameter used for book-keeping,  a small string coupling $g_s$
and a large 't Hooft coupling $\lambda=g_s N_c\sim 20$.  (Subleading  $1/\lambda$  effects of the curved geometry 
will be included only as a correction to the Pomeron intercept where small effects are important.)
The setting includes AdS$_5$-like space with a confining wall where the important number of 
transverse directions is physically identified  with \be D_\perp=3 \ee
containing the transverse plane and the holographic direction.
We refer to it as the SZ model. We note that its technical core  -- the calculation of the
Euclidean amplitude of the twisted tube exchange -- was done in~\cite{Basar:2012jb}.

The main phenomenon to be studied is the string diffusion.
At very high energies the standard large parameter \be \chi={\rm ln}(s/s_0)\gg 1 \label{eqn_chi} \ee 
 plays  the role of an {\em effective diffusion time}.

\begin{figure}[t]
  \begin{center}
  \includegraphics[width=6cm]{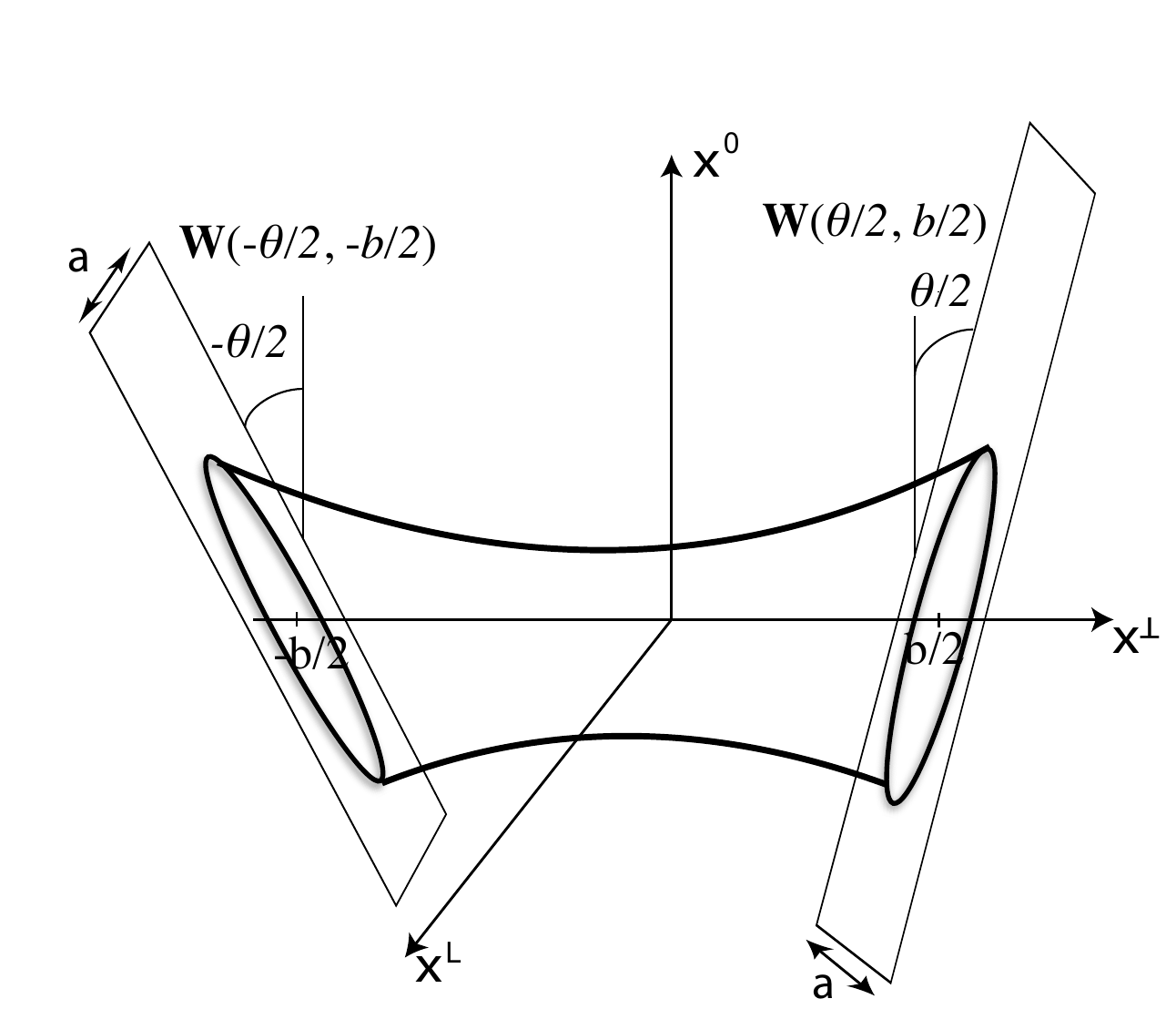}
  \caption{(Color on-line)
  Dipole-dipole scattering configuration in Euclidean space. The dipoles have size $a$ and are $b$ apart. The dipoles are tilted by $\pm \theta/2$ (Euclidean rapidity) in the longitudinal $x_0 x_L$ plane.
   }
  \label{fig_tube2}
  \end{center}
\end{figure}

 We will now review the Pomeron results  in this setting. The amplitude of the elastic dipole-dipole scattering
  reads~\cite{STOFFERS,ZAHED,Basar:2012jb}

\begin{eqnarray}
\frac1{-2is}{\cal T}(s,t; k)\approx  {g_s^2}
\int\, d^2{\bf b}\, e^{iq\cdot{\bf b}}\,{\bf K}_T(\beta, {\bf b}; k)\,\,\,
\label{4X3}
\end{eqnarray}
where  ${\bf K}_T$ is called the string (or Pomeron)  propagator.
One of its arguments, ${\bf b}$, is the impact parameter, which is the length of a ``twisted tube", 
providing a semiclassical solution to the problem. 
The other $\beta$ is the {\em circumference} of the tube. Its analogy with
the Matsubara time leads to the introduction of an  {\em effective temperature} $T$. Its value 
depends on the ``diffusion time" $\chi$ 
and is also proportional to the impact parameter  
\be \beta={1 \over T} ={2\pi{\bf b} \over \chi}  \label{eqn_T} \ee 
$\chi$ is  our large parameter (\ref{eqn_chi}). 
 The last integer argument $k$ describes the color string flux,
 known also as $N_c$-ality and related to Young tableaux of the color representations.  In particular,
 for the antisymmetric ones $k$  
  runs over  all integers till $N_c/2$ for even $N_c$, and $N_c/2-1/2$ for odd ones.
  While we will show $k$ in some formulae below, we will only use the usual string between fundamental charges (quarks)
  and $k=1$, for
 the real world  of SU(3) color. Only when we will need the large-$N_c$ counting we will recall more general groups.
 Note that the first factor in the amplitude is the string coupling  
$g_s\sim 1/N_c$, small in the standard large-$N_c$ counting.

The previous literature focused on what we now call a ``cold"   regime of the string, namely a case
 \be {\bf b} \gg \beta \gg \tilde{\beta}_H \ee 
 The former inequality follows from large collision energy (\ref{eqn_chi}) and the latter implies that
the string is nearly straight, with small effective excitations (small effective  $T$). The meaning
of the tilde on the Hagedorn temperature (or the corresponding Matsubara time $\beta=1/T$)
will be explained below in (\ref{XTHH}).
The explicit form of ${\bf K}_T$ was  calculated in \cite{Basar:2012jb} using  the Polyakov string action.

\begin{eqnarray}\label{3}
&&{\bf K}_T(\beta, {\bf b}; 1)=
\left(\frac \beta{4\pi^2{\bf b}}\right)^{D_\perp/2}  \\
&& \times e^{-\sigma\beta{\bf b}\,\left(1-(\tilde{\beta}_H/\beta)^2/2\right)}
\sum_{n=0}^\infty\,d(n)\,e^{2n \chi}\nonumber
\end{eqnarray}

The first combination of parameters in the  exponents  is  the classical action.  
Here we emphasize the length $\beta/2$ or the {\em semicircle}, which first appeared in the
semi-classical approach to pair production in an electric field process back in 1931's \cite{Sauter}.  
Note that we calculate the elastic amplitutude, in which a pair of virtually produced  open strings makes a complete circle. This amplitude is the same as  the cross section, or the modulus {\em square}
of the inelastic amplitudes, with each corresponding to a tube cut in half, or two semicircles .
Here $\sigma=\sigma_T/2$.

The  first correction in the second line  is due to the ``thermal"  excited states of the string:
it corresponds to the so called Luscher term in the string-induced potential. We wrote it 
 using the (tilde)  Hagedorn  temperature of the double string (\ref{eqn_tilde_T_H}) . While physically
 in inelastic amplitude one produces an ordinary fundamental string, the conjugated amplitude has another
 anti-string, making it into a double string.
  The last  factor contains a summation over the integer $n$ due to ``tachyonÓ string modes.
  In the Regge language those are called ``Pomeron daughter trajectories.
   Some details of the weight $d(n)$ can be found in the Appendix~\ref{sec_d(n)}.

Inserting the leading $n=0$ contribution of (\ref{3}) in (\ref{4X3}) yields the Pomeron contribution
to  the elastic dipole-dipole scattering amplitude at large $\chi$ and fixed N-ality $k$
\be
{\cal T}(s,t; k)\approx ig_s^2\,\left(\frac s{s_0}\right)^{1+\frac{k D_\perp}{12}+\frac {\alpha^\prime}{2k} t}
\label{4XX2}
\ee
Thus the resulting  Pomeron has the intercept above 1 (and corresponds to  a cross section growing with energy) 
\begin{eqnarray}
\alpha_{{\bf P},k}(0)&&=1+\frac{kD_\perp}{12}\nonumber\\&&\rightarrow 1+\frac{k D_\perp}{12}
\left(1-\frac{3(D_\perp-1)^2}{2kD_\perp\sqrt{\lambda}}\right)\nonumber\\
\label{4XX3}
\end{eqnarray}
where the first term is the Luscher
contribution and the  $1/\sqrt{\lambda}$ correction follows from the tachyonic correction (\ref{2X1}) 
in curved AdS$_5$~\cite{STOFFERS}.

   While (\ref{3}) has been derived in \cite{Basar:2012jb} from the semiclassical approach to a 
   Polyakov string, but (to leading order in $1/\lambda$) it can be alternatively derived  from
 a diffusion equation 
 \be
\left(\partial_\chi+{\bf D}_k\left({\bf M}_0^2-\nabla_{\bf b}^2\right)\right){\bf K}_T=0
\label{1}
\ee
where large $\chi$ interval is the time.   The diffusion happens in the (curved) transverse space
with the diffusion constant ${\bf D}_k=\alpha^{\prime}/2k=l_s^2/k$.
   This diffusion (\ref{1}) is nothing else but the  Gribov diffusion of the Pomeron,
leading on average to an impact parameter $\left<{\bf b}^2\right>={\bf D}_k\chi$ for close Pomeron strings. 
If the ``mother dipoles" are small in size, the diffusion is close to the UV end of the holographic coordinates and perturbative results
are expected. 
For large times or dipole sizes, ${\bf b}$ is large  and the string diffuses to the confining
holographic region  near the IR end of space, with a ``confining wall". 
The ``tachyon mass" is 
\be
{\bf M}^2_0=\frac {4D_\perp}{\alpha^\prime}\left(\sum_{n=1}^\infty \frac{n}{e^{2\chi n/k}-1}-\frac{1}{24}\right)
\label{2}
\ee
  
  The extra $z$ coordinate is different from others.
  Note that the effects of the AdS$_5$ curvature is to make it
difficult for the string to wander in the 5th dimension in the IR, effectively reducing the number of transverse
dimensions and thus the Luscher contribution.
 To account for  finite size dipoles, the string ends  are placed at 
 fixed heights $z_1,z_2$ a finite distance from the confining wall
at $z_0$. As a result, the tachyon mass 
experiences   corrections   due to the  curvature in $z$

\be
{\bf M}^2_0\rightarrow {\bf M}_0^2 +\frac{(D_\perp-1)^2}{4\alpha^\prime\sqrt{\lambda}}\,\,.
\label{2X1}
\ee
Most of the arguments to follow will be carried out for large $\lambda\gg 1$ unless indicated otherwise,
so this effect is considered small.

The sub-critical string regime discussed so far is defined by the condition $\beta=2\pi{\bf b}/\chi>\beta_H$ 
in the diffusive limit $\left<{\bf b}^2\right>=D_k\chi$. A more precise bound follows from 
the inclusion of the $1/\lambda$ corrections in the tachyon mass (\ref{2X1}) or 
\be
\beta>\sqrt{2(\alpha_{\bf P}-1)} \,\beta_H
\label{HH}
\ee
This leads to the bound  $\chi<10$ for the corrected phenomenological value of the Pomeron intercept $\alpha_{\bf P}-1=0.08$ in (\ref{4XX3}), which roughly corresponds to energies below the LHC.
This condition discriminates between a sub-critical and a critical string as we will detail below.
We note that (\ref{HH}) implies a strong coupling renormalization of the  Hagedorn temperature through the 
geometry of AdS$_5$.


\subsection{Connecting to perturbative BFKL Pomeron} \label{sec_BFKL}

The Reggeon or Pomeron as an open or closed string exchange, can be viewed as a surface
of multi-gluon exchanges. In weak coupling, the surface is dominated by rapidity ordered BFKL
ladders~\cite{BFKL}.  

The conformal nature of QCD perturbation theory as captured by the BFKL ladder
re-summation can be recovered from the close string exchange since the AdS$_5$
geometry is conformal near the boundary. 
This point can be clearly seen in the holographic construction in curved
AdS$_5$ by computing the density of wee partons ${\bf N}(\chi, z,c, r)$
(proportional to ${\bf K}_T$ in curved AdS~\cite{Stoffers:2012zw})
of small size $z$ sourced by a mother dipole 
of size $r$ in the transverse radial coordinate space  $r={\bf b}$ for fixed rapidity $\chi$.
Specifically~\cite{Stoffers:2012zw}, 

\begin{eqnarray}
&&{\bf N}(\chi, z,c, r)\approx\nonumber\\
&&
  2 \frac{e^{(\alpha_{\bf P}-1) \chi}}{\left(4\pi {\bf D} \chi\right)^{3/2}} \frac{z }{cr^2} {\rm ln}\left(\frac{r^2}{z c}\right) e^{-{\rm ln}^2\left(\frac{r^2}{z c}\right)/(4{\bf D} \chi)}  \nonumber \\
\label{WEEX}
\end{eqnarray}
The diffusion is log-normal. The analogue of (\ref{WEEX}) in the
context of onium-onium scattering was discussed in~\cite{Mueller:1993rr,Mueller:1994jq}. In particular, in the
BFKL 1-Pomeron approximation it is  given by~\cite{Salam:1995uy} 

\begin{eqnarray}
&&{\bf N}^{\bf BFKL}(\chi, z,c, r) \approx 
2 \frac{e^{(\alpha^{\bf BFKL}-1)\chi}}{\left(4\pi {\bf D}^{\scriptscriptstyle \bf BFKL} \chi \right)^{3/2}}  \nonumber\\
&&\times\frac{z}{cr^2}{\rm ln}\left(\frac{16 r^2}{zc} \right) e^{-{\rm ln}^2\left(\frac{16r^2}{zc}\right)/(4 {\bf D}^{ \bf BFKL} \ \chi)} \ ,
\nonumber
\label{DBFKL}
\end{eqnarray}
with the BFKL intercept $\alpha^{\bf BFKL}$ and diffusion constant ${\bf D}^{\scriptscriptstyle \bf BFKL}$

\begin{eqnarray}
\alpha^{\scriptscriptstyle \bf BFKL}=&&1+\frac{\lambda}{\pi^2}\,{\rm ln}\,2 \label{onium}\nonumber\\
{\bf D}^{\scriptscriptstyle \bf BFKL}=&&7 \lambda \zeta(3) /(8\pi^2)
\end{eqnarray}
Modulo the Pomeron intercept and the diffusion constant which are different (weak coupling or BFKL versus strong coupling or  holography) , the holographic result in the conformal limit is identical to the BFKL 1-Pomeron approximation.

The occurence of the $3/2$ exponent reflects on diffusion in $D_\perp=3$. This point is rather
important as it shows that the conformal nature of the QCD string is recovered if the QCD string
evolves in curved AdS$_5$ instead of flat 4-Minkowski dimensions. The curved and extra dimension
captures the dipole scale evolution or equivalently the size of the closed string exchange during the 
collision.

\subsection{Regge trajectories in SZ model} \label{sec_reggions_SZ}
For completeness, we note that Reggeon exchange with open strings can be addressed
similarly. For the Reggeon $\sigma=\sigma_T$ and the elastic scattering amplitude for
dipoles of N-ality $k$  is now

\be
{\cal T}(s,t;k)\approx ig_s^2\,\left(\frac {s_0}s\right)\,\left(\frac s{s_0}\right)^{1+\frac{k D_\perp}6+{\alpha^\prime}t}
\label{4X02}
\ee
with the extra $s_0/s$ pre-factor accounting for the normalization of the spinors traveling
on the exchanged world-sheet. This point was originally made in~\cite{Janik:2000pp} but with different
conclusions for the Reggeon intercept. At large $s$, the Pomeron exchange is dominant.
The Pomeron as a closed string can be viewed as 2 glued open strings or a pair of Reggeons
up to spin factors. As a result
the Reggeon slope is twice the Pomeron slope  while its intercept   is also twice the Pomeron intercept.

A dual description of the scattering amplitude (\ref{4X3}) is in terms of Pomerons and Reggeons in the
holographic limit. Specifically,

\begin{eqnarray}
&&{\cal T}(s,t)\approx ig_s^2(\pi a)^2
\sum_{k=1}^{[N_c/2]}
\sum_{n=0}^\infty\,\nonumber\\&&
\frac{(-1)^k}{k}\left(\frac{k\pi}{\rm{ln\,s}}\right)^{D_\perp/2-1}
d(n)\,s^{1+\frac{D_\perp}{12k}-\frac{2n}k+\frac{\alpha^\prime t}{2k}}\nonumber\\
\label{T1}
\end{eqnarray}
with all $k$ N-alities included. 
The closed string or glueball trajectories following from (\ref{T1}) are

\be
J\equiv 1+\frac{D_\perp}{12k}-\frac{(D_\perp -1)^2}{8\sqrt{\lambda}}-\frac{2n}k+\frac{\alpha^\prime }{2k}\,{\bf M}^2_{n,k}
\label{T2}
\ee
where the leading AdS$_5$ curvature correction is shown.
We note that a proper P and C parity assignment for the glueball
states follows from a Mellin transform of (\ref{T1}) and its
parity conjugate. It will not be necessary for our discussion.
For source dipoles in the fundamental representation or $k=1$,
the Pomeron trajectory corresponds to ${\bf M}_{0,1}^2$, while its
daughters to ${\bf M}_{n>0,1}^2$. Their intercepts $\alpha_{P,D}(0)$ are tied by

\be
\alpha_P(0)-\alpha_{Dn}(0)=2n
\label{T3}
\ee
while their common slopes are set by $\alpha^\prime /2$.
This is very consistent with lattice glueball Regge trajectories shown already in Fig.\ref{glueballs}.



\section{Quantum fluctuations} \label{sec_fluct}



\subsection{The temperature and the entropy} \label{sec_entropy}

   Perhaps the use of the words  ``temperature", ``entropy" etc should be in quotation marks, as  
the setting  we discuss corresponds to the QCD vacuum at
  $zero$ temperature. The reason is technical and originates from the fact that  the  exchanged strings have a world line -- membrane of a shape of a tube 
 shown in Fig.~\ref{fig_DD} --
quantized on a circle, with a periodic $\tau$ coordinate. This makes it formally identical to the thermal Matsubara formalism. Quantum string fluctuations take the form of thermal fluctuations.
The temperature is  the inverse of the tube circumference $T=1/\beta$.

\begin{figure}[t]
  \begin{center}
  \includegraphics[width=6cm]{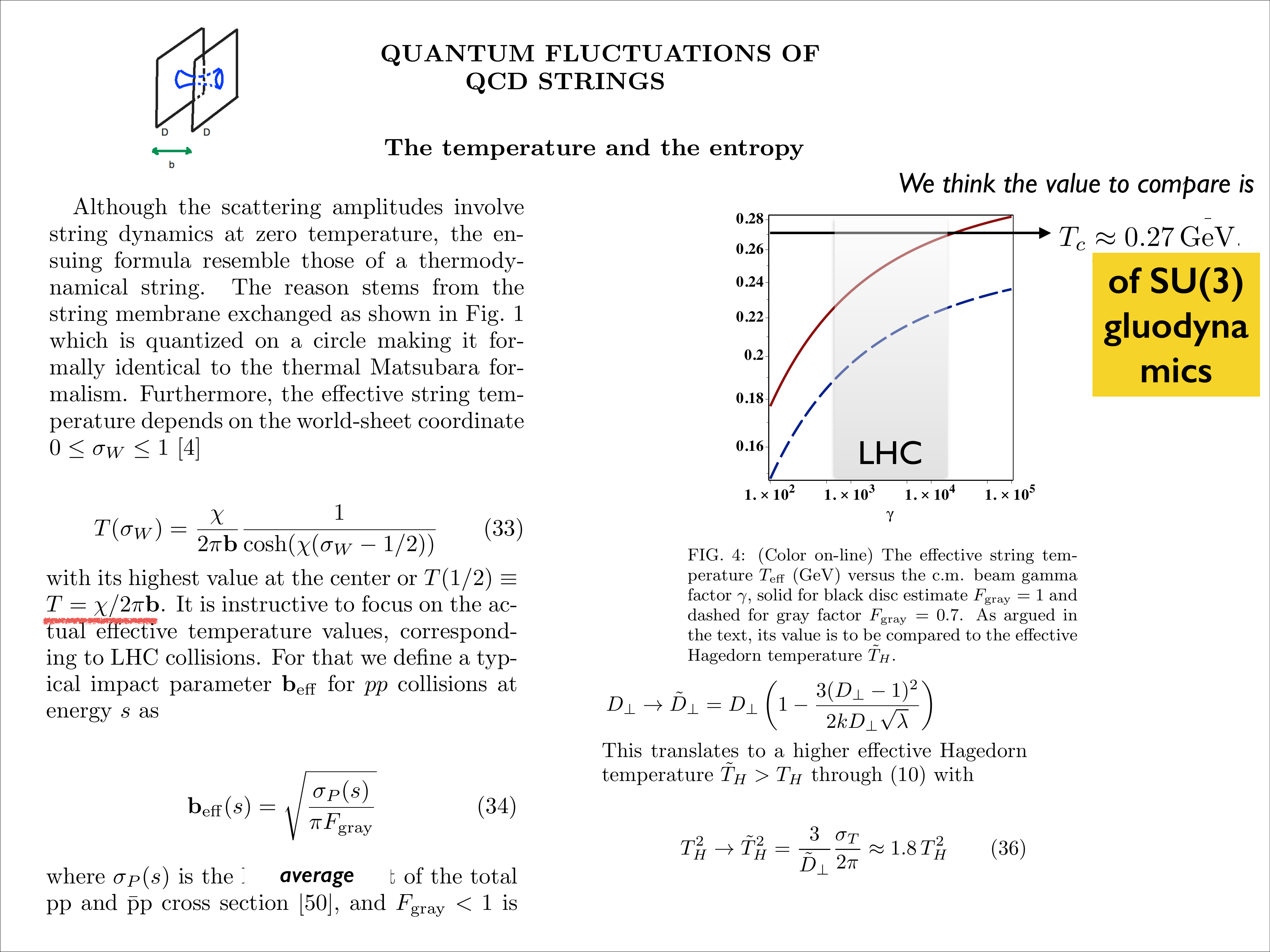}
  \caption{(Color on-line) The effective string temperature $T_{\rm eff}$ $({\rm GeV})$ versus the c.m. beam gamma factor $\gamma$.
   The curve for the
  black disc estimate $F_{\rm gray}=1$ is shown by the solid line, and for  $F_{\rm gray}=0.7$ by the dashed line. 
  The effective temperature is compared to the  critical temperature  $T_c$ of $gluodynamics$, shown by a horizontal line with an arrow, which is related with the Hagedorn temperature  by relation (\ref{THTc}).
 }
  \label{fig_Teff}
  \end{center}
\end{figure}

Furthermore, the tube circumference -- and thus the effective string temperature -- depends on the 
other world-sheet coordinate $0\leq \sigma_W\leq 1$ \cite{Basar:2012jb}

\be 
T(\sigma_W)={\chi \over 2 \pi {\bf b}} {1 \over {\rm cosh}(\chi(\sigma_W-1/2))} \label{T_versus_sigma} 
\ee
with its highest value at the center or $T(1/2)\equiv T=\chi/2\pi{\bf b}$.
It is instructive to focus on the actual effective temperature values, corresponding to LHC collisions.
For that we define a typical impact parameter ${\bf b}_{\rm eff}$ for $pp$ collisions at energy $s$ as

\be 
{\bf b}_{\rm eff}(s) =  \sqrt{\sigma_{P}(s)\over \pi F_{\rm gray}} 
\label{BEFF}
\ee
where $\sigma_{P}(s)$ is the Pomeron's part of the total pp and $\bar{\rm p}{\rm p}$ cross section~\footnote{
For definiteness, we use the PDG parameterization $\sigma_{{\bf P}}(s) =35.45+0.308*{\rm ln}^2(0.1381*\gamma^2)$ with
$s=4 M^2 \gamma^2$.}, and $F_{\rm gray}<1$ is the factor which shows how ``gray" is the nucleon.  
Inserting (\ref{BEFF}) into the effective temperature (\ref{eqn_T}) yields Fig.~\ref{fig_Teff}. 
The effective temperature slowly rises with the collision energy.
For gray or non-black-disc nucleons with $F_{\rm gray}<1$,  the effective impact parameter is larger resulting into
a  {\em downward} shift in the effective temperature.


As we noted earlier in (\ref{4XX3}) the effects of the AdS$_5$ curvature causes effectively the string
to move in effectively $\tilde{D}_\perp<D_\perp$ with

\be
D_\perp\rightarrow{\tilde{D}}_{\perp} =D_\perp\left(1-\frac{3(D_\perp-1)^2}{2kD_\perp\sqrt{\lambda}}\right)
\ee
This translates to a higher effective Hagedorn temperature $\tilde{T}_H>T_H$ through (\ref{THH}) with

 \be 
 T_H^2\rightarrow \tilde{T}_H^2={3 \over {\tilde{D}}_\perp} {\sigma_T \over 2\pi} \approx  1.8 \,T_H^2
 \label{XTHH}
 \ee
 where in the last equality we used a typical value $\lambda=20$, which gives $\tilde{T}_H\approx 0.224\, {\rm GeV}$. 
 
 The curvature-related corrections  
 shift the effective Hagedorn temperature upward.  The shift is close to the factor $\sqrt{2}$
 one expects from the double-tension gluonic strings (as discussed in the thermodynamical introduction above).
 We may argue that the higher order curvature corrections perhaps shift it a bit more, to the    
 critical temperature of the Yang-Mills theory $T_c\approx 0.27\, {\rm GeV}$ or even beyond, it, to $T_H=1.11 T_c$.
 Comparing those expectations with the effective temperature values calculated from the impact parameter in Fig.~\ref{fig_Teff}
 we find that the exchanged string is expected to reach the near-critical regime only at collision energies well above  the LHC domain.  
   This justifies that  so far most of the $pp$ collisions  are still described by a cold (far from critical) string.
( However more central collisions lead to higher $T_{eff}$ and the corresponding
 near and super-critical strings will be described in the next sections.) The thermal analogy 
 allows us to define the free energy ${\bf F}=-{\rm ln}{\bf K}_T/\beta_U$ and the entropy
corresponding to small string vibrations ~\cite{STOFFERS,ZAHED}

\begin{eqnarray}
{\bf S}=&&-D_\perp\sum_{n=1}^\infty
\left({\rm ln}\left(1-e^{-\beta_k n}\right)+\frac{\beta_k n}{e^{\beta_kn}-1}\right)\nonumber\\
&&+D_\perp\left(\frac{\beta_k}{12}-\frac 12\left(1+{\rm ln}\left(\frac{\beta_k}{2\pi}\right)\right)\right)
\label{4X1}
\end{eqnarray}
At large collision energy $\chi\gg 1$ the entropy is dominated by the last term due to the tachyon, so 
\be
{\bf S}\approx \frac{D_\perp\beta_k}{12}
\label{4X2}
\ee
Since $\beta_k=2\chi/k$ the entropy scales with the rapidity interval $\chi$. In contrast, the energy 
${\bf E}\approx \sigma{\bf b}$ with on average $\left<{\bf b}^2\right>\approx {\bf D}_k\chi$, 
scales with the root of $\chi$, and therefore is subleading for asymptotically large $\chi$. This 
is a major difference between the ``cold" regime and the others  that we will discuss below.

For clarity, let us emphasize that this entropy characterizes the number of states
of the ``tube", or strings produced at the initial virtual stage of the collision. It is obviously $not$
the number or states or entropy physically produced in the collision and observed in the detector,
although we will argue below that there is a positive correlation between the two, at least in some regimes.

\section{Near-critical strings}   \label{sec_near_critical}

So far we have discussed the so called ``minimally biased" collisions.
Their typical impact parameter was extracted from the total cross section.
Now we switch to discussing certain fluctuations in a system, corresponding to
more ``central" collisions, with the impact parameter smaller than the typical one. 
(At this point the reader may ask how experimentally one can find such an event. 
We postpone its discussion to section \ref{sec_observables} below.) 
    



\subsection{The Hagedorn phenomenon leads to string balls}

\begin{figure}[t]
  \begin{center}
  \includegraphics[width=6cm]{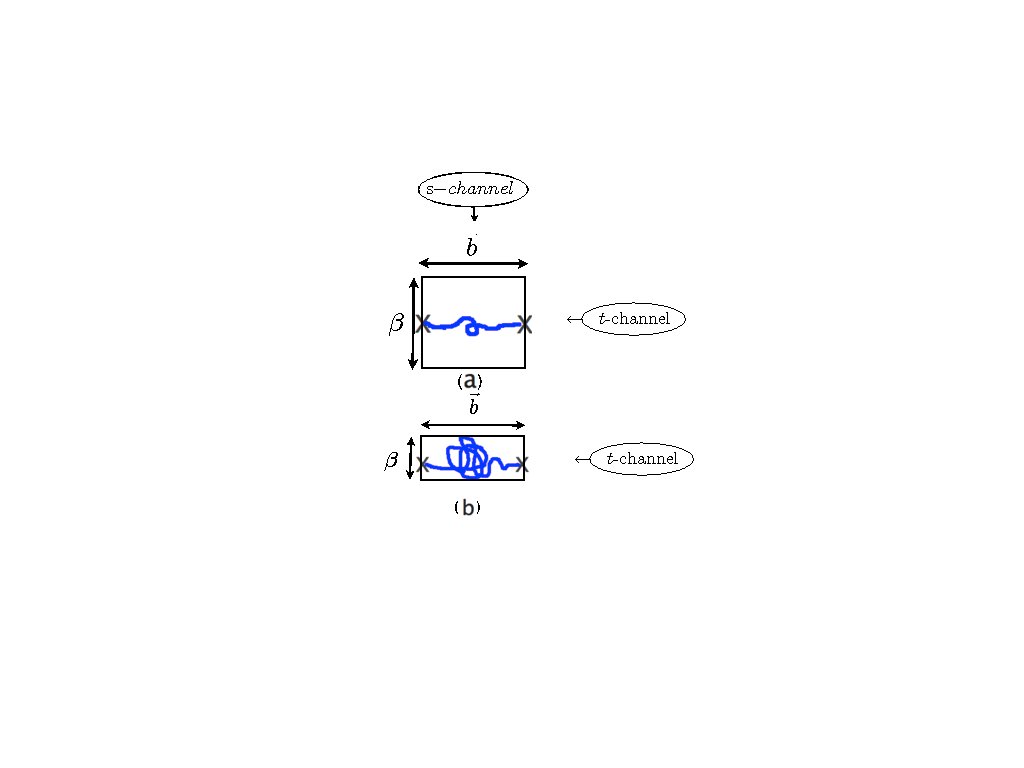}
  \caption{(Color on-line) String exchange between two sources (crosses) separated by
  the impact parameter ${\bf b}$:  the cold string case $\beta<\beta_H$ (a);  the near-critical string ball $\beta \rightarrow  \beta_H$ (b).}
  \label{SS}
  \end{center}
\end{figure}

As it is clear from the formulae given above, the smaller impact parameters
 correspond to thinner tubes and thus higher effective temperatures.
The central idea of this paper is that some radical change is expected when the effective string temperature approaches the  Hagedorn temperature $T\rightarrow  \tilde{T}_H$ (the tilde is a reminder of the curvature corrections).  The string fluctuations change from small
 as shown in~Fig.~\ref{SS}a, to large as shown in Fig.~\ref{SS}b.  The reduction of the effective string tension leads to a
 proliferation of string fluctuations.  The energy of the string and its entropy grows, as the effective 
 temperature $T$ approaches ${\tilde{T}}_H$. We will argue that in this case a string generates a massive cluster, to be called a {\em ``string ball"} below.  The physical analogy to what happens in the thermal (heat bath) setting is at the origin of this idea.

     Now, is there any connection between the effective thermodynamics of the virtual exchanged string
we discussed above, and the multiplicity of the produced hadrons?     
 The initial string configuration we discuss
 in connection with the elastic amplitude does not of course directly correspond 
 to the physical final states. Two open strings make a virtual (under the barrier) semi-circle and are then born into the physical Minkowski world as a pair of real strings thanks to the Schwinger pair production
 mechanism. Their virtual Euclidean evolution ends there. The subsequent evolution in Minkowski signature happens with
 probability one and thus is irrelevant for the scattering amplitude. It is not described by the formalism we use.  
 
  Yet, at least in the near-critical regime, one may argue that the large energy and  entropy of the string ball cluster
 is simply proportional to the physical length of the string. These strings are to be stretched longitudinally,
and then broken into pieces, corresponding to physical mesons whose multiplicity we trigger. While those
phenomena are complicated (and described by phenomenological models, e.g. those originating from 
the Lund model), we may still argue that the final multiplicity should grow with the length of the
initial but virtual string. Furthermore, we think that the final multiplicity should
simply be proportional to the initial length of the string, to its energy or entropy.

    The theoretical description of the near-critical strings can be made in the so called ``thermal scalar" formalism, suggested in \cite{thermalscalar}
(and used e.g. in \cite{Horowitz:1997jc} to be discussed in the next section). The meaning of this complex scalar field 
$\varphi$ 
is a coefficient of certain string wrapping modes with a mass 
\be m^2_\varphi={\beta^2-\beta_H^2 \over 4 \pi^2 (\alpha')^2} \label{therm_scal} \ee
vanishing at the Hagedorn point. A free field with such a mass corresponds to a free (random walk) string with 
a Gaussian diffusive distribution.   
   The description of the free string ball in the  near-critical  random walk (r.w.)
  regime is covered in detail in \cite{Horowitz:1997jc}. Let us just mention that
  its radius depends on the number of ``turns" $N$ and thus the mass as 
   \be {R_{r.w.} \over l_s} \sim \sqrt{L\over a} \sim \sqrt{ M } \label{r.w.}
   \ee
for any dimension $d$. 
 
%
%
%
%
 
\subsection{Self-interacting string balls and black holes}
   
    When any of the object gets very massive, it is amenable to a classical description.
     Sufficiently massive string balls should become black holes. 
  String theorists have studied exactly how the interpolation between the 
states of massive ``string balls" and those of black holes happen.

 Let us start with naive estimates,
which will elucidate the problem. 
The string ball can be seen as a random walk made of $M/M_s$ steps, with $M_s\sim 1/\sqrt{\alpha'}$
  the typical mass of a segment. The string entropy  is the number of segments
  \be S_{ball}\sim \frac{M}{M_s}
  \ee
  The Schwarzschild radius of a black hole in d spatial dimensions is 
  \be R_{BH}\sim \left(G_N M\right) ^{1 \over (d-2)} \label{R_BH} \ee 
   and the Bekenstein entropy
   \be S_{BH}\sim {Area \over G_N}\sim \left({M \over M_s}\right)^{d-1\over d-2} \ee
   grows with the mass as a power less than 1. Thus their equality $S_{ball}= S_{BH}$ can only be reached
   at some special  mass.  This happens when the Hawking temperature of the black hole
   is  exactly the string Hagedorn value $T_H$ and the radius is at the string scale. 
   So, at such mass a near-critical string ball
   can be identified -- at least thermo-dynamically -- with a black hole.
   
   However, in order to understand how exactly it happens one should first address the 
   following puzzle.  The random walk radius (\ref{r.w.})  does not agree with  the Schwarzschild radius $ R_{BH}$ given above (\ref{R_BH}): e.g. the former does not depend on space dimension $d$ and the latter does. So, 
something important has been missing, since a smooth interpolation
to the black hole properties  has not yet been achieved.

This goal has been reached in two steps. We believe Susskind first 
pointed out the importance of string self-gravity, and the 
consequent contraction of the ball size. Horowitz and Polchinski \cite{Horowitz:1997jc}
used mean field analysis, and  Damour and Veneziano \cite{Damour:1999aw} (whom we follow below)
completed the argument by using the correction to the ball's mass due to self-interaction. 
Their reasoning is as follows: self-interaction causes a shift in the string mass

\be {\delta M \over M} \sim  { g^2 M \over R^{d-2}} 
\ee
where $g$ is the string self coupling constant. We changed self-gravity to self-interaction because in the AdS/QCD setting
the attraction due to the scalar dilaton field is as important as gravity. (In the quoted expression above this amounts
to a coefficient change, which is suppressed anyway.) 

Omitting some technical points we proceed to
 the expression for the entropy of a self-interacting  string ball of radius $R$ and mass $M$ 
\begin{eqnarray}
S(M,R)\sim &&M\left(1-{1\over R^2}\right) \left(1-{R^2 \over M^2}\right)\nonumber\\
&&\times \left(1+{ g^2 M \over R^{d-2}}\right)
\end{eqnarray}
where all numerical constants are  suppressed for brevity.
 For  very weak coupling the last term in the last bracket can
be ignored and the entropy maximum is given by the first two terms.
This  brings us back to a random walk string ball. However, even
for very small $g$, the importance
of the last term depends not on $g$ but on $gM$. So, a very massive balls can be influenced by a very weak
self interaction (as indeed are planets and stars).
If the last term is large compared to 1, the self-interacting string balls are much smaller in size than the naive
random walk estimates suggest.

\begin{figure}[t]
  \begin{center}
  \includegraphics[width=6cm]{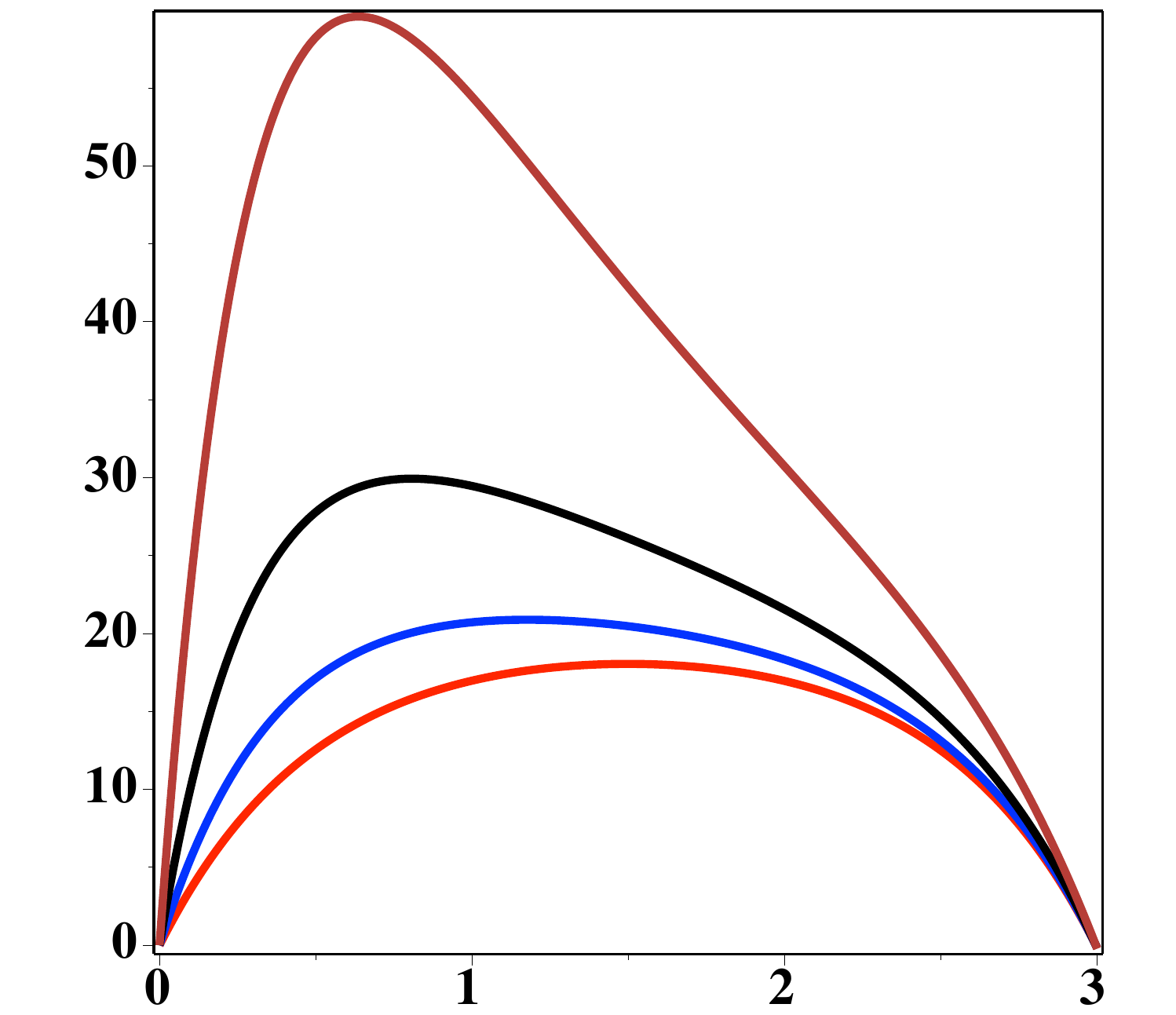}
  \caption{(Color online)  The entropy ${S}(M,R)$ as a function of ${\rm log}R$ for $M=20,d=3$.
  Four (red,blue,black and brown) curves, bottom-to-top, are for the string self coupling $g=0,0.03,0.1,0.3$.
  The corresponding shift of the maximum  is from a free string ball to a black hole.
  }
  \label{fig_self}
  \end{center}
\end{figure}

What exactly happens depends somewhat on space dimension $d$. Plots for $d=3$ (4-dimensional space-time)  
and varying coupling are shown, as examples, in Fig.~\ref{fig_self}. As one can see, a free (random-walk) string at zero
coupling has a maximum in the middle of the plot. As self-coupling grows, the string ball basically $implodes$, 
reducing its most likely radius. One can also see that it smoothly interpolates 
eventually  to the Schwarzschild radius of a
black hole. 
Numerical studies of self-interacting string balls will be reported elsewhere~\cite{KS}.

To summarize:  in the near-critical regime one finds
{\em self-interacting string balls, or $string-holes$, SH} for brevity, which 
combine a growing energy and entropy of a cluster with the implosion of its size due to self-interaction. 
 It is such objects which dominates the near-critical ``mixed phase" of QCD and scattering at intermediate
impact parameters.

The detailed consequences of this scenario for AdS/QCD models or QCD strings  remains to be worked out. In the latter case an important
ingredient of the problem is the finiteness of the scalar and tensor glueball masses.
A Yukawa-like potential would substitute to the Coulombic corrections stemming
from a massless dilaton and 
graviton of the string theory. This clearly would somewhat reduce 
the collectivity of the phenomenon. We plan to report studies of 
such string balls elsewhere.


\subsection{The scattering amplitude in the near-critical regime}\label{near-critical}

Let us now see how the  scattering amplitude and other properties of the string change as  one enters this new ``near-critical" regime.  
 Recall first the expressions discussed above, such as 
(\ref{3}), which were derived using the  Polyakov action in the regime
$\tilde{\beta}_H<\beta<{\bf b}$. They were dominated by the ground state mode $n=0$, so
\be
{\bf K}_T(\beta, {\bf b};1)\approx 
\left(\frac\beta{4\pi^2{\bf b}}\right)^{D_\perp/2}
e^{-\sigma\beta{\bf b}\,\left(1-\tilde{\beta}_H^2/2\beta^2\right)}
\label{5}
\ee
However, as the effective temperature becomes closer to the Hagedorn temperature 
$\beta\rightarrow \tilde{\beta}_H$, the string excitations are no longer small
 and the $({\tilde \beta}_H/\beta)^n$ corrections with all $n$ need to be re-summed.

The re-summed 
result  follows in the spirit of Arvis \cite{Arvis:1983fp} already mentioned and
 takes also a  square root form  (We start with the $n=0$ case, returning to other terms later.)
\be
{\bf K}_T(\beta, {\bf b}; 1)\approx 
\left(\frac\beta{4\pi^2{\bf b}}\right)^{D_\perp/2}
e^{-\sigma\beta{\bf b}\,\left(1-\tilde{\beta}_H^2/\beta^2\right)^{1/2}}
\label{6}
\ee
Clearly (\ref{6}) reduces to 
(\ref{5}) for $\tilde\beta_H/\beta\ll 1$. The first  correction is the analogue of the ``Luscher" term. This and all
other corrections have sign plus, so that each of them increase 
the amplitude, as we indicated in our sketch Fig.~\ref{fig_amplitude} near the point $A$.

Another way to say it is that
the re-summed expression
(\ref{6}) corresponds to 
 the effective string tension to vanish at  the Hagedorn point
\be
\sigma\left(1-{\tilde\beta}_H^2/\beta^2\right)^{1/2}\rightarrow 0
\label{7}
\ee
in agreement with the universal behavior observed for 
strings in a heat bath.
 As we noted above, this occurs when the impact
parameter ${\bf b}\approx \chi l_s$.

\begin{figure}[t]
  \begin{center}
  \includegraphics[width=6cm]{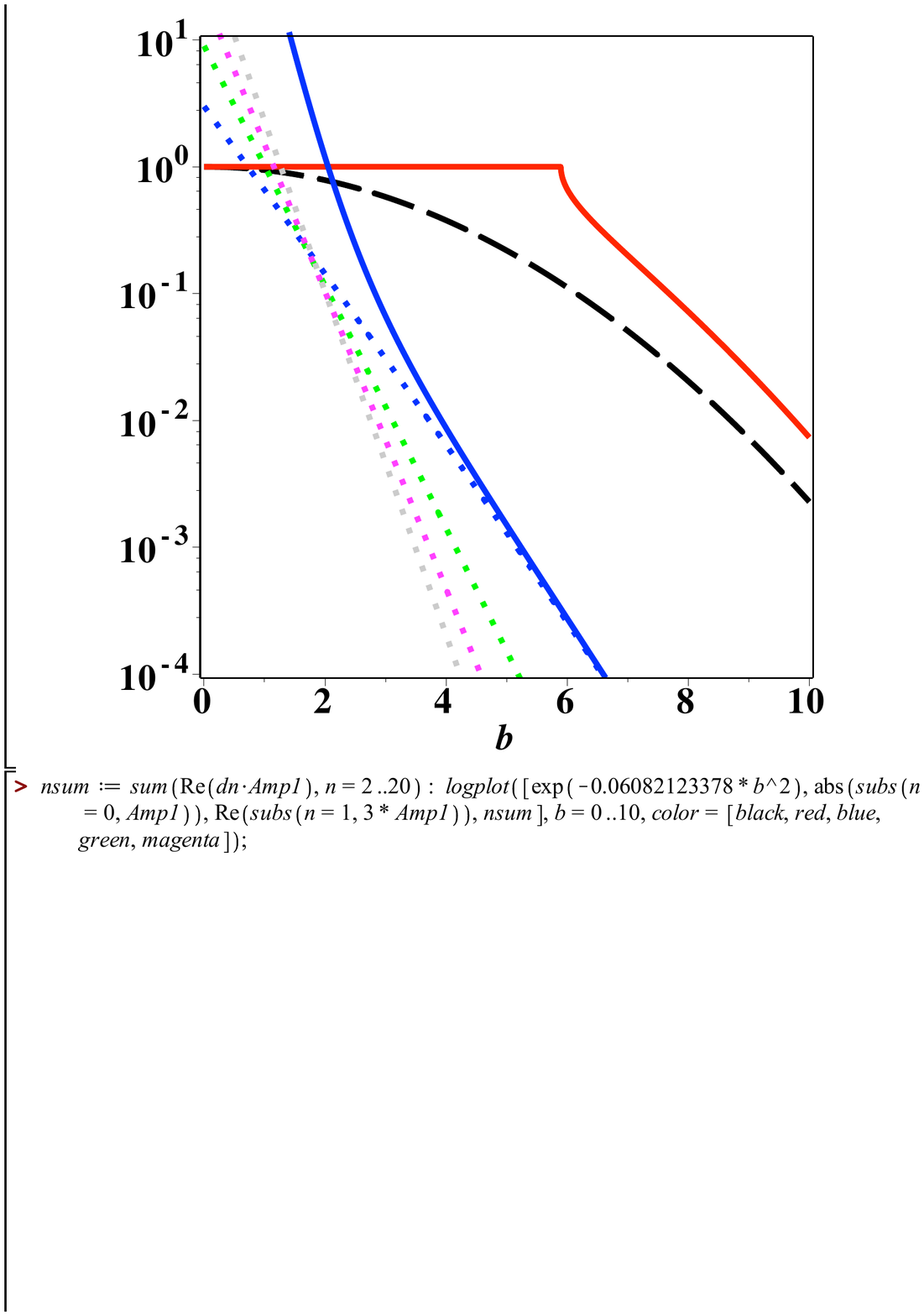}  \caption{(Color on-line) Example of $b$-dependence of scattering amplitude. The (black) dashed line
  is the original Gaussian, the (red) solid line with a kink is the re-summed version (\ref{sqrtn}), four dotted curves are 
  terms in the sum for $n=1,2,3,4$, and (blue) solid line rising to the left is the sum with exact $d(n)$.
  }
  \label{fig_sumn}
  \end{center}
\end{figure}

Now we generalize it to any $n$
\be 
\sum_{n=0} d(n) exp\left(-\sigma\beta b \sqrt{1-{\beta^2_H \over \beta^2} +{4 n \chi \over \sigma \beta b}}\right)  \label{sqrtn}
\ee
again using the idea that the term under the root is uniquely defined via its first correction known before.

Changing $e^{-n}$ to $e^{-\sqrt{n}}$ changes the convergence of the sum. Furthermore,
since at high $n$ the density of states behaves as $e^{\sqrt{n}}$  (\ref{4}), one  finds another instance of the Hagedorn
phenomenon, in which suppression of the PomeronÕs daughters will be lifted by the growing density of states $d(n)$. 
As one can see in Fig.~\ref{fig_sumn}. For smaller ${\bf b}$ the sum over $n$ diverges.
However, we should disregard this second Hagedorn point 
because it lies in the region of ${\bf b}$ to the left of the vanishing string tension for $n=0$.
Indeed in this case we are in a different super-critical phase, to be discussed below, and the expression used is not valid.



The scattering amplitude associated to such regime can be obtained by inserting
(\ref{6}) in (\ref{4X3}). The result in the saddle point approximation reads
\begin{eqnarray}
&&{\cal T}(s,t; 1)\approx ig_s^2\\
&&\left(\frac{s}{s_0}\right)^{\left(t/\sqrt{2}\right)\left(1-\frac 14(1+\sqrt{1-2/t})\right)\left(1+\sqrt{1-2/t}+1/t\right)^{1/2}}\nonumber
\label{8X8}
\end{eqnarray}
In this expression $t$ is in string units, so actually it is $ \alpha^\prime t$, and $k=1$. This expression (\ref{8X8}) reduces to the Pomeron amplitude (\ref{4XX2}) for 
$s\gg -t> 1/\alpha^\prime$. 
 One may in principle observe the corresponding modifications in the elastic scattering. However,
 we think this to be only possible at energies well above the LHC, so we will not elaborate further
 on this point. 

\section{The super-critical regime} \label{sec_supercritical}


\subsection{Strings with condensed ``thermal scalar"}

%

As we emphasized above,  at $T>T_c$ the string ball simply turns into a ball of plasma, which can be described 
in terms of deconfined colored quasiparticles, gluons and quarks. Even as we know the interaction in this matter,
known as sQGP, remains strong,
its approximate conformal symmetry requires
the pressure to be no longer subleasing, but instead jump to
 the conformal value $p=\epsilon/3$. The consequences of this fact is the ``explosive" 
 behavior to be discussed later.
  From a theoretical perspective, the simplest option is that
 the ``most central" super-critical collisions should be described via  perturbative QCD,
 e.g. by the BFKL Pomeron scattering amplitude, and thus forget about confinement and strings
 in this regime.
 
 However, one can proceed into the deconfined phase with a string-based description as well.
 An important notion, well known to string theorists, is that  a string  can be viewed as
 infinitely many fields: those are technically the coefficients of the vibrational modes. 

One approach to the supercritical region is to follow the ``thermal scalar" formalism \cite{thermalscalar}
 already discussed.  Naively, the mass square of the field $\varphi$ (\ref{therm_scal})    gets  negative. As 
 usual, it means that zero mean of that field is unstable and 
 that it develops a nonzero condensate $\left<\varphi\right>\neq 0$. Its magnitude is determined by 
 higher order terms, usually
 by the positive quartic term $ |\varphi|^4$ in the effective Landau-Ginzburg action.
 As a result, a shifted  field $\varphi - <\varphi>$ has  positive mass.
The correlator of two masses becomes of the type
\be \left<|\varphi(0)|^2\,|\varphi(x)|^2\right>=  |\left<\varphi\right>|^4+{\cal O}\left(e^{-|x| |M_\varphi|}
\right) 
\label{disconnected}
\ee 
 This phenomenon is also known as the formation of a nonzero Polyakov's  ``disorder parameter" at $T>T_c$ in finite temperature QCD.

 But a more direct and more physically appealing description is the holographic one. Since the supercritical
 regime corresponds to a trapped surface (black hole) formation, one should re-think any string-induced amplitudes:
  {\em parts of the strings inside the horizon should not be counted}. (See Fig.\ref{fig_ball} for a picture.)  This explains why
 charges are no longer connected to each other but ``liberated".  The part inside the BH cannot transmit any information outside, thus strings effectively end on the horizon: 
  so (in the leading order)
 there is no potential between the charges, as the factorized result (\ref{disconnected}) tells us.

The lesson of this section, once again, is that there is a fundamental asymmetry between the perturbative and 
the stringy  points of view. Why we don't know how to derive strings and thus the confining phase from 
a perturbative viewpoint,
one can provide a relatively simple and  logical description
of the perturbative domain starting  from  strings, even through the
deconfinement  phase transitions.

\section{The observables} \label{sec_observables}

\subsection{Elastic scattering} \label{sec_elastic}

Earlier in our discussion in the introduction we have defined a ``profile function" $F(s,{\bf b})$
in (\ref{eqn_profile}) related to the scattering amplitude as its Bessel transform. Now we tie this
to the proton wave-functions. In so far we have considered fixed size dipoles. The proton as a
quark-diquark can be viewed as a dipole. However, the dipole size fluctuates inside the proton wave-function. 
With this in mind, we now identify the curved  5th coordinate $z$ with the dipole size. 
Using the scale free coordinate  $u=-{\rm ln}(z/z_0)$ with $z_0>z$ or $u>0$, the coordinate of the 
confining wall, we define

\begin{eqnarray}
 F(s,{\bf  b})=&&\int du_1 du_2 \nonumber\\
 &&\times \,|\Psi(u_1)|^2  |\Psi(u_2)|^2  K(u_1,u_2,{\bf b},s)\nonumber\\
\end{eqnarray}

Because of the fluctuations,  the two dipole sizes are different in general. The ensuing formulae
are therefore a bit more involved. The string propagator $K$ connects 2 points in the curved
 3-dimensional transverse space, say  (-{\bf b}/2, $u_1$),(+{\bf b}/2, $u_2$). Much like bulk 
 propagators, $K$ can be simply expressed in terms of combination of arguments involving
 the ``chordal distance" $\xi$ in curved AdS between these 2 points. Specifically, 
 
\be {\rm cosh}(\xi)={\rm cosh}(u_2-u_1)+{1 \over 2}{{\bf b}^2 \over  R_{ADS}^2} e^{u_1+u_2} \ee
where $R_{ADS}$ is the radius of the effective space (in ${\rm GeV}^{-1}$ and similarly for ${\bf b}$).
Since the AdS space is walled at $z_0$, there is a reflected propagator. The invariant 
``chordal distance" $\xi^*$ is set by the image and reads

\be 
{\rm cosh}(\xi_*)={\rm cosh}(u_2+u_1)+{1 \over 2}{{\bf b}^2 \over  R_{ADS}^2} e^{u_2-u_1} 
\ee
The string amplitude derived in \cite{Stoffers:2012ai} indeed  takes a more intuitive diffusive 
form in such variables

\begin{eqnarray}\label{Kcurved}
 K(u_1,u_2,{\bf b},s)=&&{g_s^2 \over 4} (2\pi \alphaÕ)^{3/2}\nonumber\\
 &&\times \left( \Delta(\chi,\xi) +  e^{2u_1} \Delta(\chi,\xi_*) \right)\nonumber\\
\end{eqnarray}
\be 
\Delta(\chi,\xi)=\frac{{\rm exp}{[-{\xi^2 \over 4 D \chi}+\chi(\alpha_P-1)] }}{4\pi D \chi}   {\xi \over {\rm sinh}(\xi)} 
\ee 
with $D=1/2\sqrt{\lambda}$ and $\alpha_P-1=1/4$. This corresponds to a ``tube amplitude" with small excitations,
that is to be applicable at very large ${\bf b}$.

For intermediate ${\bf b}$, we follow
the arguments in section \ref{near-critical}  and generalize (\ref{Kcurved}) to the near-critical
regime by the Arvis-style substitution of the first two terms to full square root containing all higher order Nambu 
string corrections 
\begin{eqnarray}
  -{\xi^2 \over 4 D \chi}  +\chi(\alpha_P-1) \rightarrow - {\xi^2 \over 4 D \chi} 
\left[ 1- {\tilde{\xi}^2 \over \xi^2} \right]^{1/2}     \label{chi_profile} 
\label{SUB}
  \end{eqnarray}
where $\tilde{\xi}=\chi \sqrt{8D(\alpha_P-1)}$.

\begin{figure}[t]
  \begin{center}
  \includegraphics[width=6.5cm]{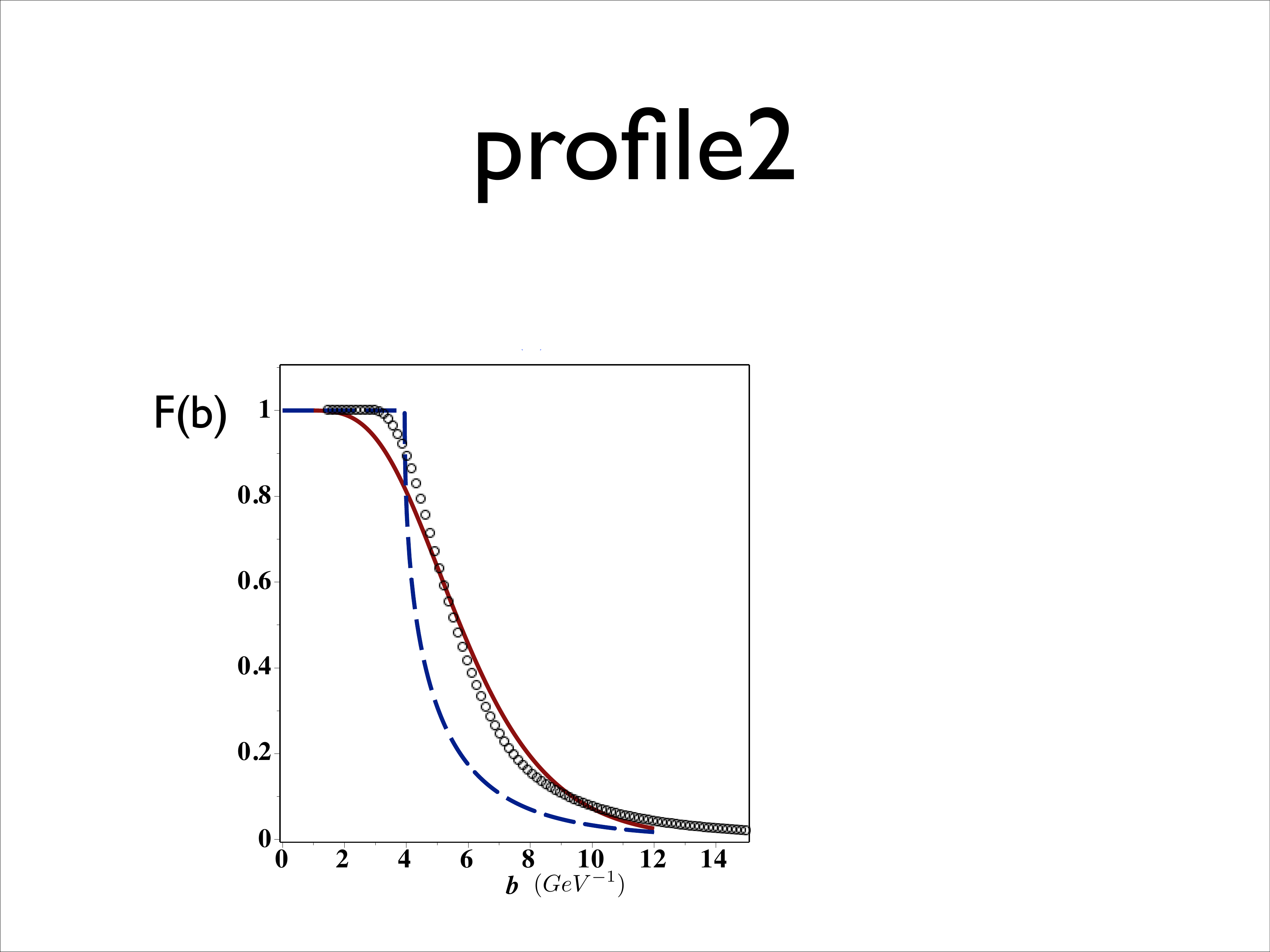}
    \includegraphics[width=6cm]{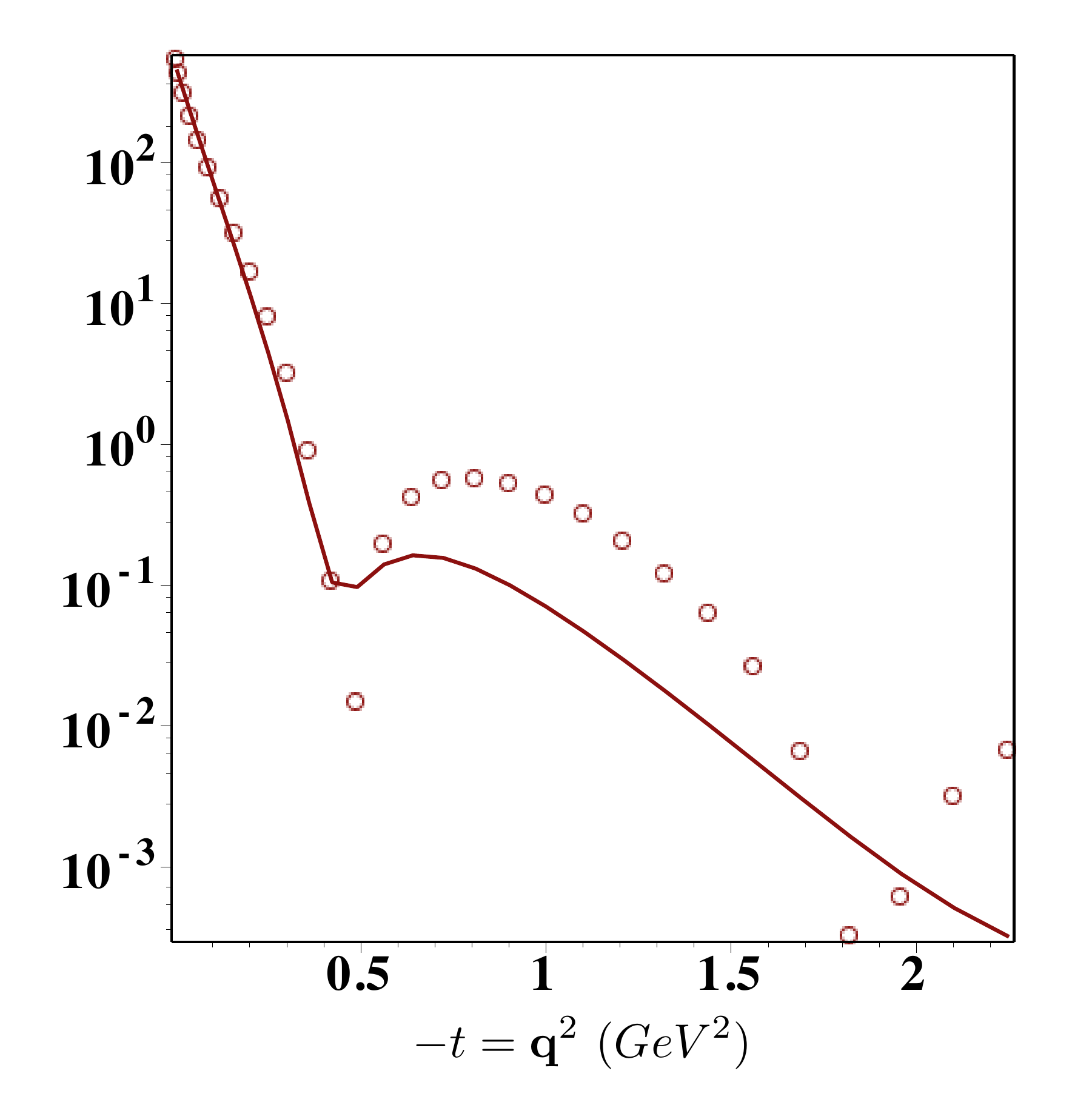}
  \caption{(Color on-line) The profile function $F({\bf b})$ versus the impact parameter ${\bf b}$ is shown
  in the upper plot for LHC $\sqrt{s}=7 \, {\rm TeV}$ energy.
  The solid line is the same curve as in Fig.\ref{fig_profile} corresponding to the BSW data parametrization.
  The dashed line is the shape corresponding to the approximation (\ref{chi_profile}) for fixed
  sizes of the dipoles $u_1=u_2$, while the circles correspond to the profile with the
  fluctuating dipoles. The lower plot shows the corresponding absolute value squared of its Bessel transform
  as a function of momentum transfer. 
  }
  \label{fig_profile2}
  \end{center}
\end{figure}

To streamline the numerical analysis of the profile function in the near-critical regime,
we now make some bold simplifications:  (i)  ignore the reflection term;
(ii) include the distance-independent amplitude at small $\xi$ in the super-critical regime;
(iii)   fix the overall   normalization constant in such a way that at the point of the vanishing square root 
$K= 1$. This results in a relatively simple expression

 \be  K(u_1,u_2,{\bf b},s)\approx e^{- {\xi^2 \over 4 D \chi} \,
{\rm Re}\,\sqrt{1- {\tilde{\xi}^2 \over \xi^2} }} \label{shape} \ee 
If the dipole sizes $u_1,u_2$ are fixed and equal,
  the profile has the shape  shown in Fig.~\ref{fig_profile2} by the dashed line.   
Note the singularity corresponding to the end of the intermediate regime and the beginning of the black hole formation
(called  in some previous  plots point $B$). Such a singularity -- or a jump in the function following the first order transition
in string thermodynamics -- in the scattering profile would not be phenomenologically acceptable. Its Bessel
transform would generate a too small power of $t$ in the  differential cross section $d\sigma/dt$ at large $t$.

However, it is  expected on general grounds (and also known experimentally from diffraction) that nucleons are 
strongly fluctuating, from one event to the other. In our approach the nucleon is simplified to a color dipole
(between a valence quark and a diquark). Its fluctuations are described by the wave function in the 5-th dimension 
$\Psi(u)$.
Making various shapes and widths of this function and performing the averaging over the string endpoints  $u_1,u_2$
we got profiles in between the dashed line (at small fluctuations) to the shape shown by circles in Fig.~\ref{fig_profile2}, for large 
${\cal O}(1)$ fluctuations of the dipole sizes.
While the resulting profile is rather close to the BSW parameterization of the data,  
the modulus squared of its Bessel transform (shown in the lower part of  Fig.\ref{fig_profile2}) 
show more visible differences. The dips in particularly are much more pronounced.
The reason is that our model contains only the  imaginary amplitude, while the BSW data parametrization
has the real part as well.  
Since this paper is about qualitative effects, we have not tried to make more sophisticated shapes of the
string propagator which would fit the $d\sigma/dt$  TOTEM data better.  See also~\cite{Stoffers:2012ai} for a similar
fit using fixed dipole wave-functions. The shape (\ref{shape}) is of course a caricature, with a  square root
 singularity. All we want to emphasize is that it corresponds to the end of the Hagedorn transition
 and approximately describes the structure seen in the elastic amplitude profile.
 
\subsection{The final state of inelastic collisions} \label{sec_inelastic}

   Stricktly speaking, this subject goes beyond the content of the present paper, as we have only analyzed the Euclidean
   part of the system path.    Still we would like to make some general comments.
    
The perturbative approach to the Pomeron, based  on re-summing gluon ladders, was studied both at the level of the elastic and inelastic amplitudes. Feynman diagrams  can be ``cut" by the  well known unitarity rules,  predicting  single-gluon and two-gluon distributions in the inelastic  collisions. However, at small $|t|$ we cannot justify  perturbative methods.  While the use of 
strong coupling $\lambda$  and large $N_c$ yield  ``fishnet diagrams" resembling a string world-sheet, 
the correspondence was never made sufficiently precise.  

Our approach uses from the start a string description (strong-coupling).
The elastic amplitude, in particular, was calculated using an
under-the-barrier  ``tube", virtual string exchange,  resulting in the ``holographic Pomeron" described above.
In principle, we could have followed the system, from its Euclidean birth to its Minkowski evolution, 
and calculated the string configurations, all the way to their final breaking and hadronization. 
We plan do to so elsewhere.

Nevertheless, we would like to speculate on this issue, arguing that some properties of the virtual string should find their
way to observable final states. 
As it is well known from experiment,  final hadrons -- mostly pions -- come from certain clusters,
hadronic resonances. Those are 
well described by the Lund-type model,  including  string breaking into certain segments,
before final decays into pions.
Our conjecture is that in the high multiplicity events associated with ``string balls" as we detailed above,
these clusters are perhaps $larger$. 

%

\begin{figure}[t]
\includegraphics[width=7cm]{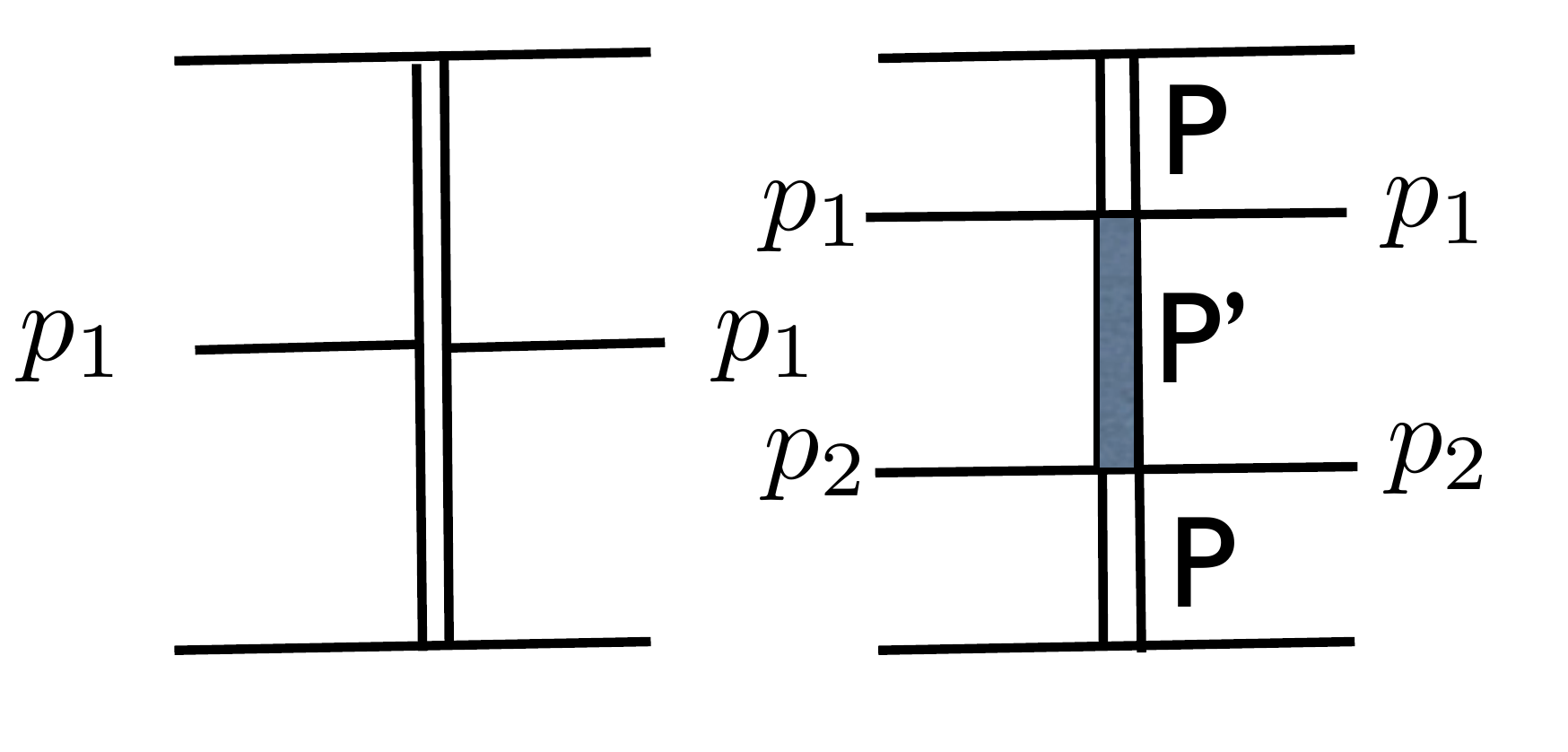}
  \caption{(Color on-line) Mueller-Kancheli diagrams for single and double particle production from Pomeron exchange.
  In Fig.2 the shaded region indicates the excited Pomeron $P'$. }
  \label{KM}
\end{figure}

In standard Regge phenomenology one uses the so called Kancheli-Muller diagrams ~\cite{MK}, see 
 Fig.~\ref{KM}, to calculate the single and many-hadron spectra. We focus now on the two-particle correlations.
  From  the $t$-channel point of view, (nearly) unclustered two-particle spectrum
  corresponds to the  Pomeron exchange, and further
    clustering corresponds to ``daughters" of the Pomeron
   with  $n>0$ excitations.   
The lines in  Fig.~\ref{KM}
are the corresponding propagators, which we do know.
They naturally satisfy the usual relations, in which a propagator can be written as a convolution of two propagators, 
integrated over the intermediate points. So we attempt now to use those, in the spirit of Kancheli-Mueller rules,
in an attempt to describe clustering. Including the leading  Pomeron and its first daughters to the 2-particle correlations, one expects the following
rapidity dependence
\be 
{dN \over d \Delta y}= C_P e^{\Delta y(1-\alpha_P(0))}+C_{P'} e^{\Delta y(-\Delta\alpha_1)} +\ldots \label{eqn_PP'}
\ee
The second contribution stands for the first  daughter, while the dots for the higher daughters. 
Note that the Pomeron has an empirical intercept of  $1.08-1.20$, making the first contribution
slightly rising with the rapidity interval, as is indeed observed. The  ``Pomeron daughter" contribution  rapidly
decreases with the rapidity interval since the difference of intercepts is large
(\ref{daut1}).

  \begin{figure}[t]
\includegraphics[width=8cm]{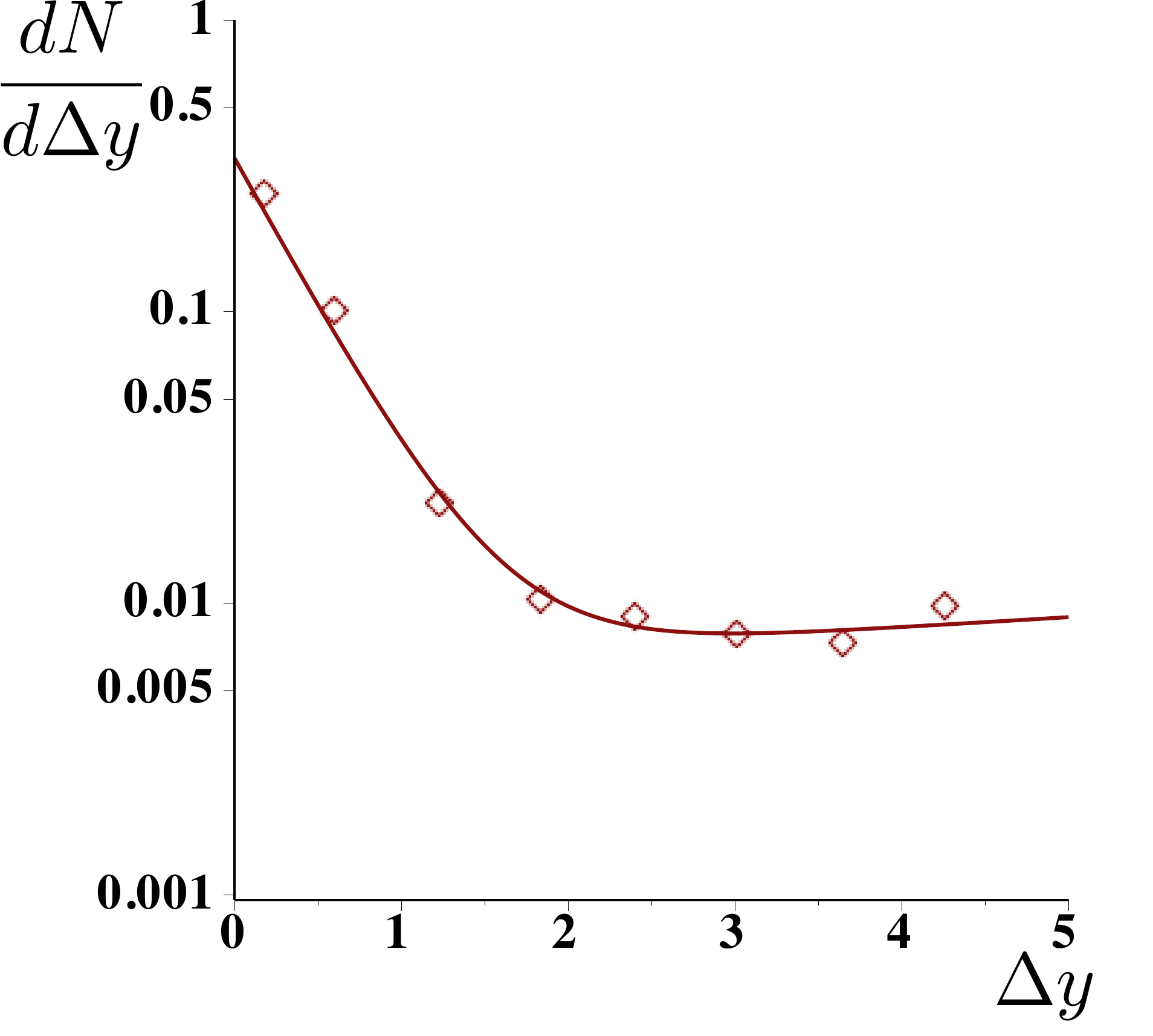}
  \caption{(Color online) Two particle correlation function fitted to $P$ and $P'$ exchanges (\ref{eqn_PP'}). The points are from the CMS data, Fig~2 of \cite{Velicanu:2011dp}, for the high multiplicity bin $N>110$ and $2<p^{trig}_\perp<3\, GeV, 1<p^{assos}_\perp<2\, GeV$. }
  \label{corr}
\end{figure}

The two particle correlation functions in the high multiplicity events are measured by the CMS
collaboration, although publicly available data are quite limited. 
The peak at small $\Delta \eta$ is usually interpreted as ``jet-generatedÓ. We doubt this to be the case 
since it is well seen at small $p_t\sim 1-3\, {\rm GeV}$ where jet contribution is small.  

Somewhat surprisingly, the  approach based on $t$-channel exchanges works well, even for high
multiplicity events.
In Fig.~\ref{corr} we show the experimental data on the two-particle rapidity distribution from CMS which we 
fitted as a function of $\Delta y$.  The fit suggests $\Delta\alpha_1=2.2\pm 0.2$, which is in the vicinity of the 
value (\ref{daut1}) obtained by the Regge extrapolation of the lattice glueball data. 
The most notable feature of this fit is the fact that the coefficient of the ``Pomeron daughter"
is larger by a factor $\sim 30$ than that of the leading Pomeron. We take it as the first direct
confirmation of a large cluster production in high multiplicity events. 
Such a strong enhancement of the subleading Pomeron is also supported by our
holographic estimate~(\ref{T3}). To summarize this point,  we suggest that
 the so-called ``jet-peak"  structure seen in two-particle correlators, is actually a hadronic cluster 
 originating from a string-ball. Its dependence on multiplicity should be studied systematically.

   This suggestion should of course be tested further. The peak is
 not only in the rapidity  $\Delta y$ variable, but it also has a certain shape 
in azimuthal angle  $\Delta\phi$. At this time, we have not analyzed whether this shape
can or cannot be described by the exchanged Pomeron and its daughters. 

Another prominent observed structure is  the so called away-side
peak at $\Delta\phi\approx \pi$. At large $p_\perp$ this is ascribed to di-jet events.
At smaller  $p_\perp$  the away-side balances kinematically the trigger particle. 
 If the Pomeron is described perturbatively, via gluon ladders
in weak coupling, then the back-to-back correlations are natural. In central 
$pA$ collisions those are   enhanced, and a quantitative discussion of this effect is available 
due to Dusling and Venugopalan \cite{KEVIN}. 
As shown by those and other authors, gluon diagrams also generate certain elliptic asymmetry $v_2$,
as the  impact parameter direction is dynamically different from the other transverse direction.

  Let us at the end of this section suggest another 
  interpretation, based on our view of a cluster -- remnant of the string ball produced at the initial time. 
  As it is clear  from Fig.~\ref{fig_ball}, the tension of the string
ends pulled along both beam directions should supply the ball with angular momentum $J$
normal to the beam and the impact parameter. Since $J$ is conserved in the cluster decays,
the products would carry it away, making the distribution anisotropic in azimuth (enriched in the impact parameter 
direction). 
  
\subsection{Explosions of the super-critical fireball}

We already mentioned the pressure of the QGP $p\approx (1/3) \epsilon$. 
 A very important consequence of this pressure is that the fireball explodes.  While the details of 
 such phenomenon are already well studied via  heavy ion (AA) collisions, it still came as a surprise to many
that   in pp and pA collisions  very high multiplicity  events  
can indeed display  collective explosion. The now famous ``ridge"
 is now interpreted as a hydrodynamical elliptic flow. Higher azimuthal harmonics of the flow also have been
 studied, and they are also surprisingly well described by hydrodynamics.
 
 However, azimuthal harmonics  are just  relatively small deformations of the hydro flow: the main effect is the so called radial flow (in the transverse plane).
  It has been predicted in  version 1 of our paper \cite{Shuryak:2013ke} that the  
  radial  flow in pp and pA should  {\em exceed in magnitude} that previously observed in AA collisions.
 Very recently this prediction has been confirmed by CMS and ALICE collaborations, as they have measured the spectra of identified secondaries  ($\pi,K,p,\Lambda$ etc), see version 2 of that paper (now in PRC).

A relatively small size of the fireball produced at freezeout,  together with a strong radial flow, 
leads to  a conclusion that it must start with an {\em extremely small and dense initial
state}.  This correlates well with the attractive string self interaction and its tendency to implode
(dual to gravitational collapse). Thus the explosion that follows may look puzzling.
The answear is that it is not the string ball which explodes, but the Hawking radiation created by its 
horizon. Naturally it only happens if it is {\em massive enough}. So  one expects to find a certain threshold
in multiplcity. 

   Unfortunately 
   an estimate of the corresponding critical multiplicity (above which an explosion should happen) is not so straightforward.
We can calculate the entropy of the string, as it leaves the Euclidean (under-the-barrier) part of its 
path, but we do not describe its further evolution in the Minkowski world, including its fragmentation
into the observed hadrons. While this evolution does not change the probability of the process, it 
generates  new entropy.  
   
   Let us suggest two arguments to this point.   The first
   argument is a lower bound. Since the entropy never decreases, the string-ball entropy
  should provide a lower bound on the final entropy and thus the multiplicity. 
The critical multiplicity ${\bf N_c}$ associated with the explosion of the black-hole is  limited by
entropy near the Hagedorn temperature

\be
{\bf N_c} > 7.5\,{\bf S}\approx \sigma\beta_H{\bf b}
\left(1-\tilde{\beta}_H^2/\beta^2\right)^{-1/2}
\label{19X1}
\ee
where the conversion factor of 7.5 is borrowed from the entropy-to-hadron density relation at freeze out.
Substituting $\tilde\beta_H/\beta-1\approx 1/N_c$ in (\ref{19X1}), we find
the critical charge particle multiplicity to be
\be
{\bf N_C} > 7.5\,\chi/2\approx 50
\label{19X2}
\ee 
 This bound 
(\ref{19X2}) surprisingly agrees with
 the measured threshold charge tracks multiplicity of
$N> 50$ for events with
the ridge, according to  the CMS collaboration
\cite{Abelev:2012cya}. The agreement is however purely accidental. If one includes 
the actual acceptance of the CMS detector  in $p_\perp$, as well as include a factor 1.5 for neutral secondaries,
the actual multiplicity is larger than the CMS track count  by about a factor of 3.

The second argument is that since string fragmentation is essentially a local process, 
the resulting multiplicity is proportional to the string length. Since in the near-critical
regime the string entropy is proportional to its length as well, we suggest that the
string entropy we calculate and the final multiplicity should be, in this regime, proportional
to each other.

\section{The string ball as an effective black hole} \label{sec_BH}


  This section contains some additional theoretical material,
which is not necessary for the overall understanding of the rest of the paper.
However, it provides
interesting alternative physical analogies and results.  
In it we will   show that the ``string ball" state is thermodynamically (and perhaps 
in other respects)  
dual to a black-hole (BH).

This trapped surface is very different from that appearing in the
effective thermal BH AdS metric used in most of the thermal
AdS/CFT studies of the sQGP. The horizon is $not$ static, and is $not$ placed only along the holographic 
z-direction  (the SZ model has a holographic
 z-coordinate   with a confining wall at $z=z_0$, not a thermal horizon). The BH is dynamically produced , as schematically shown in Fig.~\ref{fig_ball}.  Here we will focus on  the part of its horizon normal to  the $longitudinal$  L-direction

The near-critical string has a propagator (\ref{6}) that behaves like a thermal ensemble with Unruh
temperature $1/\beta_U$. Its free energy
or pressure ${\bf F}=-{\rm ln}{\bf K}_T/\beta_U$ ~\cite{STOFFERS} is small
\be
{\bf F}(\beta, {\bf b}) \approx k\sigma\,{\bf b}\left({1-\frac{\tilde{\beta}_H^2}{\beta^2}}\right)^{1/2}
\label{9}
\ee
but its energy
and  entropy are  large
\begin{eqnarray}
&& {\bf E}=\partial_{\beta_U}(\beta_U k {\bf F})\approx k\sigma{\bf b}\left({1-\frac{\tilde{\beta}_H^2}{\beta^2}}\right)^{-1/2}\\
&& {\bf S}=\beta_U^2\partial_{\beta_U}{\bf F} \approx (\tilde{\beta}_H^2/\beta)k\sigma{\bf b}\left({1-\frac{\tilde{\beta}_H^2}{\beta^2}}\right)^{-1/2}\nonumber
\label{10}
\end{eqnarray}
For $\beta\approx \tilde{\beta}_H$ this coincides with the first law of thermodynamics for black-holes in Rindler coordinates 
as noted by Susskind~\cite{SUSSKIND}

\be
{\bf S}\approx {\beta_H}{\bf E}=2\pi\,({\bf E}l_s)
\label{11}
\ee
and vanishingly small pressure (\ref{9}). We  note the Rindler temperature $T_R=1/2\pi$ and therefore the Rindler energy ${\bf E}_R={\bf E}l_s$. 
The emergence of a Rindler temperature is expected since the stringy Pomeron exchange is characterized by a line 
element~\cite{STOFFERS}
\be
ds^2\approx -a^2\,\rho^2dt^2+d\rho^2+ds^2_\perp
\label{11X}
\ee
with a Rindler acceleration $a=\chi/{\bf b}$.  At this regime the acceleration is
$a=k/l_s$.  On the streched horizon  at
$\rho=l_s/k$ in (\ref{11X}), the warping of time is 1 since ${t}/{t_\rho}=({b}/{\chi})/{\rho}\rightarrow {l_s}/k{\rho}$.
A cartoon of the string ball as a black-hole is shown in Fig.~\ref{fig_ball}.

The transverse area of the black-hole is the area of the diffusing string in rapidity
\be
A_{BH}=2\pi^2\left(\sqrt{\chi/k}\, l_s\right)^3
\label{11XX1}
\ee
in transverse $D_\perp=3$ provided that the diffusion length  in the z-direction is 
within the confining wall. As a result, we have the Bekenstein-Hawking type relation

\be
\frac{ {\bf S}_{BH}}{A_{BH}}\equiv \frac 1{4G_5}
\label{12}
\ee
with an effective  Newton constant  

\be
G_5=\pi^2 \left((\chi/k^3)(1-\tilde{\beta}_H^2/\beta^2)\right)^{1/2}\,l_s^3
\label{13}
\ee
For a fundamental string, the Planck and string constants are related with $G_5$ through $G_5=l_P^3=g_s^2l_s^3$.
We recall that in the large $N_c$ counting $g_s^2\approx 1/N_c^2$. 

The transmutation of the near-critical strings into a black-hole at the string 
scale was foreseen by Susskind and others in the context of string-based gravity~\cite{SUSSKIND,MANY}. 
Furthermore, it was later shown that the Bekenstein-Hawking
formulae emerge from a direct  statistical counting of quantum string states.
In hadronic collisions at  large rapidity
$\chi$, the effective relation  (\ref{13}) shows  that this transmutation  can be achieved in a twofold way: (i) one
is discussed in this paper
$\beta/\tilde\beta_H\rightarrow 1$ (the near-critical regime); and another (ii) is a more exotic case possible in large  $N_c$
limit, namely exchange of a string with very large color charge
 $k/\chi\rightarrow \infty$ which we do not discuss.

Empirical estimates based on DIS data 
analysis ~\cite{STOFFERS} suggests that the saturation scale is $z_0\approx 2/{\rm GeV}$, 
so that the diffusion length is far from the confining wall for 
$\sqrt{\chi/k}l_s<z_0$ or $\chi<16$ for $k=1$. For very high energy collisions, however, given by $\chi>16$ 
the diffusion length reaches the confining wall. This should modify
scattering at super high energies, in particularly the transverse area
(\ref{11XX1}) is now changed to
\be
A_{BH}\approx 2\pi^2z_0\left(\sqrt{\chi/k}\, l_s\right)^2
\label{11XX2}
\ee
with the corresponding changes in the effective Newton constant estimate 

\be
G_5=(\pi^2/k) \left(1-{\tilde\beta}_H^2/\beta^2\right)^{1/2}\,(z_0l_s^2)
\label{13X}
\ee

\subsection{Dual derivation of the string propagators} 

%
%

We can explicitly check that 
the tachyon thermodynamics (\ref{9}) and (\ref{10}) follows from the large $n$ excitation spectrum
of the NG string by using the modular transformation and the saddle point approximation in
flat space.
The modular transform of the transverse string propagator is an exchange $b \leftrightarrow \beta $,
which corresponds to going into the close string description from the open strings.
It is basically a change of coordinates in string quantization,  describing the same ``tube" configuration. 
Indeed, the modular transform of (\ref{3})
can be cast as
\begin{eqnarray}\label{XB1}
&&{\bf K}_T(\beta, {\bf b}; k)\approx \\
&&\sum_{n=0}^\infty\,d(n)\,e^{-\sigma\beta{\bf b}\,\left(1-{\bf b_c}^2/{\bf b}^2+2\pi n/\sigma{\bf b}^2\right)^{1/2}}\nonumber
\end{eqnarray}
with ${\bf b_c}=\left((\pi D_\perp)/(12\sigma)\right)^{1/2}\equiv \pi l_s$ and the density of states
(\ref{4}). The NG form has been subsumed. (\ref{XB1}) is seen to diverge for $\beta\leq \beta_H$. The divergence is controlled by a large
$n$ saddle point, 

\be
n_S\approx \frac{\sigma{\bf b}^2}{2\pi}\frac 1{(\beta/\beta_H)^2-1}\gg 1
\label{XB2}
\ee
for which (\ref{XB1}) is to exponential accuracy

\be
{\bf K}_T(\beta, {\bf b}; k)\approx {e^{-\sigma{\bf b}\sqrt{\beta^2-\beta_H^2}}}
\label{XB3}
\ee
in agreement with the tachyon result above.

The string energy at the large $n$ saddle point (\ref{XB2}) is

\be
{\bf E}\approx \sigma {\bf b}\sqrt{\frac{2\pi n_S}{\sigma {\bf b}^2}}=\frac{\sigma{\bf b}}{\sqrt{\beta^2/\beta_H^2-1}}
\label{XB4}
\ee
and the corresponding entropy is

\be
{\bf S}\equiv {\rm ln}\,d(n_S)=2\pi\sqrt{\frac{D_\perp n_S}6}-\frac {D_\perp}4{\rm ln}\,n_S
\label{XB5}
\ee
which is seen to satisfy the zero pressure condition ${\bf S}\approx \beta_H{\bf E}$ in leading order.
They are the tachyonic energy and entropy in the Hagedorn limit discussed above. This is
expected since the modular transform allows us to cross from the $\beta<{\bf b}$ 
regime of long and close strings, to the $\beta>{\bf b}$ of short and open strings. The 
two descriptions match at the border ${\bf b}\approx \beta$.

At the Hagedorn limit, a long and 
space filling string, with $D_\perp$ dimensions,  is a very efficient way to 
carry  large entropy. 
The analogy between a string ball and black hole thermodynamics shows
that in fact it carries the largest entropy density possible!
 With this in mind and for simplicity, consider a Polyakov string made of $D_\perp$ harmonic oscillators
immersed in a heat bath with finite but large Rindler temperature $1/\beta_R$. The  energy of the string is
dominated by the high frequency modes,

\begin{eqnarray}
{\bf E}_R\approx {D_\perp}\sum_{n=1}^\infty \frac{n}{e^{\beta_{R}n}-1}
\label{16}
\end{eqnarray}
For large $1/\beta_R$ it is black-body

\be
{\bf E}_R\approx \frac{\pi^2}{2\beta_R^2}
\frac{D_\perp}{3}
\label{16X1}
\ee
Through the first law of thermodynamics (\ref{11}) we can enforce the zero pressure condition
on this highly excited string, with

\be
{\bf S}\equiv {\bf S}_R\approx \beta_R{\bf E}_R=\frac{\pi^2}{2\beta_R}\frac{D_\perp}3
\label{16X2}
\ee

\subsection{Viscosity at the Rindler horizon} \label{sec_horizon}

Viscosity can be defined via certain limits of the correlators of the stress tensor, known as the Kubo formula.
Thus one does not need hydrodynamics to calculate it, just the stress tensor.
To assess the primordial viscosity, we follow \cite{ZAHED} and write the needed expression on the streched horizon
for the excited string
\be
{\eta_R}=\lim_{\omega_R\to0}\frac{A_R}{2\omega_R}\int_0^\infty d\tau e^{i\omega_R\tau}{\bf R}_{23,23}(\tau)
\label{16X22}
\ee
with $A_R$ the area of the black-hole and $\tau$ a dimensionless Rindler time. The retarded commutator of the 
normal ordered transverse stress tensor for the Polyakov string on the Rindler horizon reads

\be
{\bf R}_{23,23}(\tau)=\left<\left[T_\perp^{23}(\tau), T_\perp^{23}(0)\right]\right>
\label{16X3}
\ee
with 

\be
T^{23}_\perp (\tau) =\frac{1}{2A_R}
\sum_{n\neq 0}\,:{a_n^2}{a_n^3}:\,e^{-2i n\tau}
\label{16X4}
\ee
and the canonical rules $\left[a_m^i,a_n^j\right]=m\delta_{m+n,0}\delta^{ij}$. The averaging in (\ref{16X3}) 
is carried using the black-body spectrum as in (\ref{16}). The result is

\be
\eta_R=\lim_{\omega_R\to0}\frac{A_R}{2\omega_R}\frac{\pi}{2A^2_R}\frac{(\omega_R/2)^2}{e^{\beta_R\omega_R/2}-1}
=\frac 1{A_R}\frac{\pi}{8\beta_R}
\label{16X5}
\ee
We note the occurence of the Bekenstein-Hawking or Rindler temperature $\beta_{BH}=\beta_R$ in the thermal
factor.


Combining (\ref{16X2}) for the entropy to (\ref{16X5})   yields the viscosity
on the streched horizon

\be
\frac{\eta_R}{{\bf S_R}/A_R}=\frac 1{4\pi}\left(\frac 3{D_\perp}\right)\equiv \frac{1}{4\pi}
\label{18}
\ee
which, for $D_\perp=3$, is precisely the celebrated universal value from AdS/CFT. The result
 (\ref{18}) is remarkable as it follows solely from a string moving at large
``time" $\chi$ in non-critical dimensions but near its Rindler horizon, not in transverse coordinate $z$.
It emerges naturally in  the near-Hagedorn regime.

The result (\ref{18}) for the critical Pomeron as a close string exchange on the streched horizon for large $1/\beta_R$
is to be contrasted to the same viscosity ratio but for the low-$T$  Pomeron as a close string exchange far from the horizon
for small $1/\beta_k$~\cite{ZAHED}
\be
\frac{\eta_\perp}{{\bf S}/A_\perp}=\frac 12 \frac 1{4\pi}\left(\frac {2\pi }{\beta_k}\right)^2 \left(\frac{3}{D_\perp}\right)
\label{19}
\ee
The ratio is small at large rapidity.
(\ref{19}) reduces to (\ref{18}) for $\beta_k\rightarrow \beta_R$ up to a factor of $1/2$, showing the
non-commutativity of the two limits. Indeed, for small $1/\beta_k$ the non-critical Pomeron is described by the Polyakov 
action whereby the zero pressure condition (emblematic of a near-Hagedorn or black-hole in Rindler coordinates)
does not hold. 

Furthermore, the
relation (\ref{16X5}) yields  an effective viscosity for finite
frequency (but still zero momentum) to be thermally supressed for the large frequency modes $\omega_R$. 
\be
\eta_R(\omega_R) =\frac{\pi}{16A_R}\frac{\omega_R}{e^{\omega_R/2T_R}-1}
\label{16XX5}
\ee

The on-set of the black-hole is followed by  Hawking radiation of string bits of frequency $\omega_R/2$
as is explicit in (\ref{16XX5}) and stressed further below. In particular, for finite wavenumber $k_R$ in Rindler
units, the suppression is physically expected to follow from the substitution

\be
\omega_R\rightarrow \sqrt{\omega_R^2+k_R^2}
\label{16XXX5}
\ee
and therefore exponential as well. The effective viscosity $\eta_R(\omega_R,k_R)$ at higher gradients -- larger $k_R$-- would indeed imply a smaller effective viscosity in pp than in AA. This point is similar to the Lublinsky-Shuryak re-summation scheme~\cite{Lublinsky:2009kv}.

Concluding this discussion of the viscosity let us make the following comment. While one can use the Kubo formula for any
setting in which the stress tensor is defined, the resulting viscosity itself is of hardly any use outside of  hydrodynamics. As we emphasized above, phenomenology indicates that in an ``explosive"
regime with very high multiplicity there are hydrodynamical flows. Alas,
 both for the cool subcritical strings and the near-critical strings flows are absent.
The results of this subsection can only be used for the near-critical regime. Specifically,
they can either be used to account for non-hydro dissipative phenomena, or perhaps even
for the viscosity at the late stages of the explosive process, as the system returns to the near-critical regime.

\subsection{Hawking radiation} \label{sec_Hawking}
 
In a typical pp and pA collision in the ``cold" regime, a pair of strings is created in the scattering 
process and then stretched longitudinally to finally decay via the Schwinger pair-production 
mechanism. The decay process is captured by the Lund model in event generators.
The production of the final -- observables -- entropy and temperature in the ``near-critical" regime 
are related to its black-hole-based description. 
Standard particle emission from a black hole is described as the Hawking radiation. 
 
 We ascribed to high-multiplicity events 
  a somewhat different particle emission mechanism. This emission is fully thermal. However, it does not require
long equilibration time of the fireball, and it takes place because the near-horizon zero point oscillations
of quantum fields  apparently appear in a thermal form.  One may call it ``prompt thermal emission",
 not delayed by the usual equilibration processes. At this point 
our  approach is similar in spirit but different in details to the Unruh-Hawking effect discussed
 in \cite{DIMA}. 

The power spectrum or Hawking  emission per unit time from a black-hole is generic. 
For our rapidly moving string it involves a black-hole in $1+4$ dimensions
with the extra dimension accounting for changes in the dipole scales. In 
$D_\perp+2$ dimensions 
it reads~\cite{SCALAR}

\begin{eqnarray}
&&d^{D_\perp+1}{\bf P}=\\
&&\sum_{s}\sigma_s(\omega)\frac{\omega}{e^{\omega/T_{BH}}+(-1)^{2s+1}}
\frac{d^{D_\perp+1}k}{(2\pi)^{D_\perp+1}}\nonumber
\label{20}
\end{eqnarray}
We have only kept the dominant S-wave contributions. Here $T_{BH}=T_R=1/(2\pi l_s)$.
The sum runs over the spin $s$ of the emitted particle with $\sigma_s(\omega)$ the 
S-wave absorption cross section or grey-body factor of a spin-s on a black-hole. For
$\omega l_s\ll 1$ 

\be
\sigma_s(\omega)\approx \kappa_s\,A_{BH} \equiv 4\kappa_sl_P^{D_\perp +1}\,{\bf S}
\label{21}
\ee
with $A_{BH}$ the area of the black-hole. The last identity follows from
the Bekenstein-Hawking type relation and shows that the power spectrum
is extensive with the entropy. For scalars $\kappa_s=1$~\cite{SCALAR}. 
As the Hawking emission through (\ref{20}) unfolds, the mass and radius of the 
black-hole decreases, causing the Hawking temperature $T_{BH}$ to increase.
The emission process is inherently a non-equilibrium one. Here and for simplicity
we assume it to be quasi-adiabatic with (\ref{20}) adjusting to the change in $T_{BH}$.

For massless particles $\omega=|k|$ in (\ref{20}). The luminosity defined as 
${\bf L}(\omega)=d{\bf P}/d\omega$ for $D_\perp=3$ is

\be
{\bf L}(\omega)=\frac{A_{BH}}{8\pi^2}\sum_s\kappa_s
\frac{\omega^4}{e^{\omega/T_{BH}}+(-1)^{2s+1}}
\label{22}
\ee
It is a black-body spectrum from a 5-dimensional space where the black-hole
originated from. 

%
%

As many of these black-holes are expected to be released in AA collisions they
are the seeds of the primordial matter viewed as a collection of these tiny black holes.
Primordial Hawking emission of partonic constituents as well as electromagnetic radiation
is what current heavy ion colliders are probing. We recall that for $\chi<16$ we have
$A_{BH}\approx \chi^{3/2}$ while for $\chi>16$ we have $A_{BH}\approx \chi$
because of confinement in the holographic or conformal direction of the string. Therefore,
we estimate the tresholds ${\bf N}_T(\chi)$ for the large multiplicity events with 
explosive hydrodynamical flow to scale with beam rapidity as

\ba
\frac{{\bf N}_T(\chi_1)}{{\bf N}_T(\chi_2)}=\left(\frac{\chi_1}{\chi_2}\right)^{3/2}  
\label{24}
\ea
for $\chi_{1,2}<16$ and 1 for $\chi_{1,2}>16$, 
irrespective of whether it is pp, pA and AA collisions.

 \section{ Discussion} \label{sec_discussion}
 \subsection{Summary} \label{sec_summary}

We have started by a review of  the SZ Pomeron model, based on  an exchange of
 a non-critical string  in curved
AdS$_5$-like space with a confining wall. For typical collision events at current energies, including the LHC domain,  the Pomeron follows from the string quantized via the scalar Polyakov
action for the slightly excited string oscillators. A relatively small Luscher term generates the intercept of the Pomeron  (\ref{4XX2}),
which for $D_\perp=3$ and, with  a finite $1/\lambda$ correction,  yields a value acceptably close to the phenomenological 
soft Pomeron intercept. The slope and the ``daughter" trajectories are also found to be at the phenomenologically appropriate
places.

In this paper,  we further discussed  fluctuations  of the virtual strings, describing those by an effective temperature.
For typical min-bias collisions we found $T_{\rm eff}$ to be sufficiently far from the Hagedorn temperature $\tilde{T}_H$,
to justify the use of the ``cold" regime for the SZ Pomeron and its excitations. 
(Recall that tilde is a reminder of the up-ward shift in the temperature caused by the AdS$_5$ curvature).

The essential part of our work is about either higher-than-LHC collision energy or about
 more central collisions, with an impact parameter ${\bf b}$ less than typical. These collisions  have higher $T_{\rm eff}$ which approaches $\tilde{T}_H$ the Hagedorn temperature. We argued that in this case the string enters a new {\em near-critical
 regime}, in which  one expects the proliferation of long strings in the form of a  self-interacting ``string ball". Such phenomenon at the  corresponding temperature in truly thermodynamical setting is well known, but we argued that it should also
 happen without a heat bath, with an individual string created in the collisions.
 
 We further argued that as the mass density of the string ball reaches a sufficiently high value, the string ball becomes a black hole.
 At still lower impact parameter
  the transition to
 the third -- post-critical or
 explosive -- regime takes place, in which the system becomes amenable to a macroscopic -- hydrodynamical --description. 
It is in this regime that  strong radial, elliptic (the so called ``ridge") and even triangular flows have been detected. 
We argued that the second regime would get dominant at the energies corresponding to the highest end
of the LHC energy domain.


   While these phenomena do not (yet) correspond to the typical (min-bias) collisions  at existing experimental conditions,
   being still in the ``cold string" regime and amenable to the SZ Pomeron description developed earlier,
   a certain fraction of the {\em more central} events  should display the newly
   suggested regimes.  We argued that   the        
high multiplicity pp and pA events, triggered experimentally by certain criteria, are dominated by such 
regimes. In particular, we suggested that the  production of  a ``string ball" cluster in 
the middle of the string (mid-rapidity) is the reason for this multiplicity.


   The theoretical description of the new regimes is as follows. 
When the effective  temperature  approaches the Hagedorn temperature,  string excitations are no longer small, and
 the expression for the  string propagator (\ref{3}) is to be reconsidered. We do so  by 
using the known results for the re-summed confining potential with all-order Luscher terms
for the Nambu-Goto string action,  resulting in the  new expression (\ref{5}). For such a 
string its  tension effectively vanishes, leading to a
{\em ``string ball"} formation.

   All properties of a sufficiently massive  string ball are 
 shown  to reach those of a black-hole. 
 The particle production from such a string-ball follows Hawking thermal radiation pattern. 
 Unlike most holographic models,  this black hole does not have a horizon along the $z$ direction.
 It is produced in the collision and  its Rindler horizon  is along
the longitudinal direction. This black-hole is 5-dimensional, with 3 transverse coordinates, 2 spatial
ones and 1 conformal $z$ describing the scale evolution. The black-hole radius and area are set by the 
Gribov diffusion length, which grows with the collision energy $\chi={\rm ln}(s/s_0)$ as $\chi^{1/2}$, and 
$\chi^{3/2}$ for $\chi<16$, respectively. For very high collision energies $\chi>16$ the area growth is reduced to $\chi$ because of confinement along the holographic or conformal direction of the string.

  We have thus argued that for sufficiently central collisions the final state should contain  remnants of the string-hole. 
While we have not discussed in this paper its evolution after $t=0$ (the moment when strings appear from under-the-barrier),
let us add at least two comments. These remnants should be seen as clusters visible in two- (or more)-body rapidity correlations.
Furthermore, as evident from Fig.\ref{fig_ball},  pulling the strings along two arrows longitudinally would provide
the cluster an angular momentum. Its magnitude would only be limited by the string breaking. This means that
the produced clusters should have angular momentum $\vec{J}$,  that maybe significant and normal to the scattering plane.
This momentum would generate certain angular correlations in the transverse plane.   

 Using the Kubo formula for string excitations, we found that on
the stretched or Rindler horizon the  shear viscosity to entropy ratio is precisely $1/4\pi$,
for $D_\perp=3$.  It is the same as for the AdS/CFT black hole, in spite of the fact that
these two black holes are very different. Ours is dynamical with a horizon normal to the longitudinal
coordinate, while  the AdS/CFT one is static, with a  horizon normal to the transverse holographic
z-direction. 

We have further argued that when the temperature exceeds the Hagedorn value,
 one approaches the post-critical  regime. Its most adequate description
is generation of a black hole.
 Its Hawking radiation is seen as  a QGP fireball. As a result,  the transition to the deconfined phase  unleashes
a large pressure with $p\approx \epsilon/3$ and the stringy black-hole explodes hydrodynamically, 
following the general scaling of viscous hydrodynamics in small volumes.  The macroscopic treatment of 
these effects  is discussed elsewhere \cite{Shuryak:2013ke}.

\vskip 1cm
{\bf Acknowledgements.}
We would like to thank  Alex Stoffers, Gokce Basar, Dima Kharzeev and Derek Teaney for discussions.
This work was supported in part by the U.S. Department of Energy under Contract No. DE-FG-88ER40388.

\vskip 1.5cm

\begin{appendix}
\section{On the density of states} \label{sec_d(n)}
There are many definitions for the string density of states.
As noted in~\cite{HV},  in mathematics it goes back at least to 1918
 in the famed Hardy-Ramanujan paper. One definition consists in expanding 
 the string of products
 
\be \prod_{k=1}^\infty \left({1 \over 1-\xi^k}\right)^{D_\perp}=\sum_{n=0}^\infty d(n) \xi^n \ee
For our case $D_\perp=3$ and the first
coefficients are 1, 3, 9... as easily obtained by an expansion in series.

\begin{figure}[t]
  \begin{center}
  \includegraphics[width=6cm]{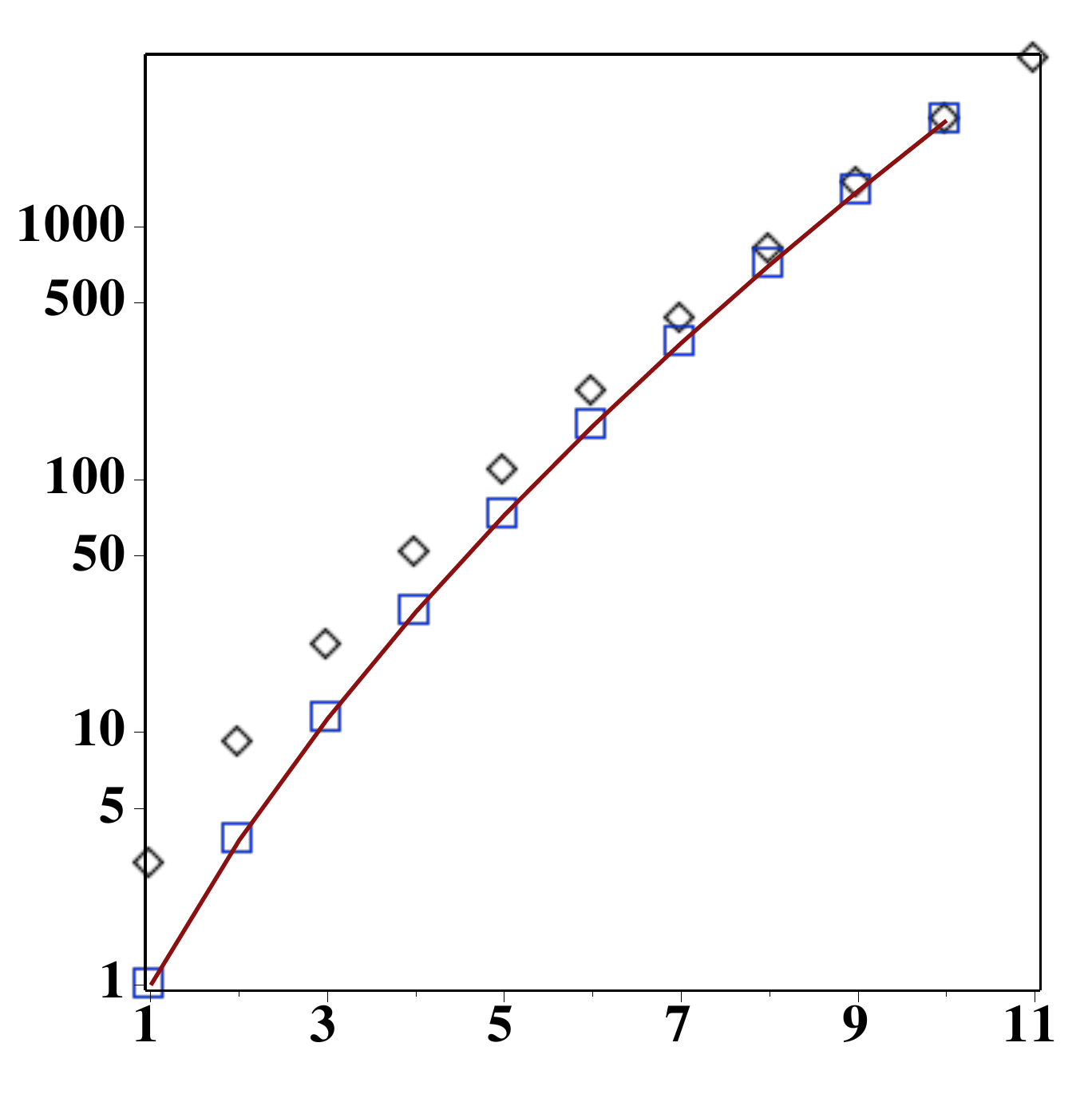}
  \caption{(Color on-line) The density of states $d(n)$ versus $n$. Black diamonds show the
  exact result, while the squares and the curve
  correspond to the asymptotic expression (\ref{4}).
  }
  \label{fig_d_n}
  \end{center}
\end{figure}

  The asymptotic density of states  is known, see e.g. \cite{DGV}  
\be
d(n\gg 1)\approx C e^{2\pi\sqrt{D_\perp\,n/6}}/n^{D_\perp/4}\,
\label{4}
\ee
where $C$ is a constant. We have checked the validity of this
 formula for the first dozen terms, see Fig.~\ref{fig_d_n}.
 We have chosen to normalize exactly the 10th term, using $C=0.01174701111$. 

\end{appendix}

\end{document}